\documentclass[pdflatex,sn-mathphys-num]{sn-jnl}

\usepackage{graphicx}
\usepackage{amsmath,amssymb,amsfonts}
\usepackage{booktabs}
\usepackage{array}
\usepackage{multirow}
\usepackage{xcolor}
\usepackage{textcomp}
\usepackage{url}
\usepackage{bm}

\newcommand{\mainfigurewidth}{0.96\textwidth}

\newcommand{\R}{\mathbf{R}}
\newcommand{\Z}{\mathbf{Z}}

\newcommand{\angstrom}{\,\text{\AA}}

\begin{document}

\title[Prior-guided adaptive refinement]{Prior-guided adaptive local refinement of foundation interatomic potentials}

\author[1]{\fnm{Yunlong} \sur{Wang}}\equalcont{These authors contributed equally to this work.}
\author[2]{\fnm{Jianzhen} \sur{Ma}}\equalcont{These authors contributed equally to this work.}
\author[1]{\fnm{Yusu} \sur{Zhang}}
\author[1]{\fnm{Jiahang} \sur{Li}}
\author[1]{\fnm{Qinqiu} \sur{Cheng}}
\author[1]{\fnm{Junjie} \sur{Yuan}}
\author[1]{\fnm{Wenrui} \sur{Yan}}
\author[3]{\fnm{Yucheng} \sur{Zhang}}
\author[3]{\fnm{Chen} \sur{Yu}}
\author*[2]{\fnm{Weibin} \sur{Chu}}\email{wbchu@fudan.edu.cn}
\author*[1]{\fnm{Hongxia} \sur{Hao}}\email{haohongxia@pjlab.org.cn}
\author*[1]{\fnm{Jielan} \sur{Li}}\email{lijielan@pjlab.org.cn}
\affil[1]{\orgname{Shanghai Artificial Intelligence Laboratory}, \city{Shanghai}, \postcode{200232}, \country{P. R. China}}
\affil[2]{\orgname{Key Laboratory for Computational Physical Sciences (MOE), State Key Laboratory of Surface Physics, Institute of Computational Physical Sciences and Department of Physics, Fudan University}, \postcode{200433}, \city{Shanghai}, \country{P. R. China}}
\affil[3]{\orgname{Hangzhou Research Center, Huawei Technologies Co., Ltd.}, \postcode{310052}, \city{Hangzhou}, \country{P. R. China}}

\abstract{Foundation interatomic potentials enable simulations across
diverse materials, but task-specific inaccuracies can require additional
first-principles data. Obtaining these data is challenging when the
relevant local environments arise in large simulation cells.
Here we introduce the Prior-guided Adaptive Refinement Scheme (PARS),
which identifies weakly supported environments during molecular dynamics
by comparing their frozen representations with prototypes of pretraining
environments, without requiring initial task labels or model ensembles.
Selected environments are reconstructed as compact periodic cells for
first-principles labelling and iterative refinement.
Applications to phase stability, surface migration and vacancy
clustering show improvements in target properties, including an 81\%
reduction in the mean absolute error of Ti vacancy-cluster energies
per vacancy relative to the pretrained model.
For Si/Ge, a foundation potential refined using 800 labelled cells
containing 10--128 atoms is subsequently distilled into a compact
potential without additional first-principles training labels.
In simulations containing approximately 362,880 atoms,
the distilled model predicts an interfacial thermal resistance
consistent with that of a potential trained directly on
density-functional theory data.
These results demonstrate a route for adapting foundation potentials
to target applications through local reference calculations.}

\maketitle

\section{Introduction}\label{sec:introduction}
Machine-learned interatomic potentials (MLIPs) approximate the
density-functional theory (DFT) potential-energy surface at a cost that permits
long molecular dynamics (MD), and derive conservative forces from a learned
energy\cite{behler2007,bartok2010,drautz2019}. Equivariant graph models have
improved their data efficiency\cite{batzner2022,batatia2022mace}, and
foundation MLIPs trained on large materials datasets now provide a broad
starting point for new
tasks\cite{smith2017ani,chen2022m3gnet,deng2023chgnet,yang2025mattersim,batatia2025foundation,barros2026omat24}.
Broad coverage does not guarantee task accuracy. Universal potentials
systematically soften surface, defect, migration-barrier and high-energy
configurations\cite{deng2025softening}; frozen-transfer studies still require
task-specific data\cite{radova2025}; and ionic and interfacial systems can
depend on long-range interactions that local architectures
omit\cite{zhang2022dplr,cheng2025les,mishin2021}. Refining a foundation
potential raises two practical questions: which task environments need
additional first-principles labels, and how can those labels be obtained
at an affordable computational cost?

Active learning (AL) addresses data acquisition through selective reference
calculations and model updates. D-optimality extrapolation grades for
moment-tensor potentials, ensemble deviations in DP-GEN, Gaussian-process
variances in FLARE and distances in a model's own latent space provide
different indicators for selecting new training
data\cite{podryabinkin2017,zhang2019dpgen,zhang2020dpgen,vandermause2020flare,janet2019latent,tan2023uq}.
Batch, bias-driven and multiscale variants extend this
strategy\cite{wilson2022,kulichenko2023udd,vandermause2022hpt,erhard2024}. Recent workflows extend active learning to pretrained potentials.
DP-EVA combines shallow-ensemble uncertainty with sampling in
pretrained descriptor space to select fine-tuning
data\cite{liu2026dp}.
Bayesian on-the-fly fine-tuning instead incorporates pretrained
parameters into a prior and uses predictive uncertainty to
trigger reference calculations during
MD\cite{rensmeyer2026fly}.
Three considerations motivate the design of a
foundation-potential refinement workflow.
First, acquisition criteria derived from the evolving model
change during refinement. Although this adaptation is useful
for tracking what the current model has learned, it does not
provide a fixed measure of support relative to the original
pretraining distribution.
Second, constructing informative uncertainty estimates for
pretrained potentials requires additional choices.
Shared pretrained initialisation can limit the diversity of
fine-tuned ensembles\cite{chtchelkatchev2026}, while readout
ensembles and quantile estimates provide alternative approaches
to uncertainty assessment\cite{bilbrey2025uq}.
A fixed pretraining-support reference offers a complementary
diagnostic without requiring initial task labels or an ensemble.
Third, the environments that need labels often arise in cells
too large for DFT. Local reference calculations have been
enabled by amorphous-matrix embedding in the Si--O
system\cite{erhard2024}. IDEAL embeds high-uncertainty
substructures into low-uncertainty
environments\cite{kong2023ideal}.
The reconstruction procedure and its computational cost
remain important design choices when converting selected
local environments into compact periodic cells for
first-principles labelling.

The computational challenge extends beyond labelling. Many properties that motivate
refinement require length and time scales far beyond DFT and costly even for a
foundation model. The thermal boundary resistance (TBR)
of a semiconductor heterointerface is a demanding example. It is obtained from
non-equilibrium MD over tens of nanometres and nanoseconds, and it depends on
how the phonon spectra on the two sides of the interface overlap. Empirical
potentials tend to stiffen phonon dispersions and underestimate the TBR of
Si/Ge interfaces relative to time-domain thermoreflectance
measurements\cite{yue2025cpl,cheng2021tdtr}. Compact MLIPs such as
neuroevolution potentials (NEPs) are fast enough for such
simulations\cite{fan2021nep,fan2022gpumd,song2024nep4}, but they must be
trained on data that cover the relevant environments. Distillation offers a
route: a foundation model adapted with a limited set of DFT reference data generates labels
on which a compact model is trained\cite{gardner2025distill}, and compact
models can also be distilled by matching energy Hessians\cite{amin2025hessian}.
The distilled potential can only learn what the labelling model describes, which
returns the problem to the first question: which DFT labels are needed
to improve the foundation potential for the target application.

Pretrained representations have already been used for data
selection. DIRECT combines pretrained descriptors with
dimensionality reduction and stratified sampling to obtain
representative training structures\cite{qi2024robust}.
More recently, activation and neural tangent kernels derived
from a pretrained MACE model have been used as acquisition
signals for MLIP fine-tuning\cite{varga2026pretrained}.
A frozen pretrained foundation model offers a reference that does not move. Its
representation already organises the environments seen in pretraining.
Studies of foundation-model representations show that their latent geometry
contains information about chemical similarity and atypical
structures\cite{li2026platonic}. A frozen copy of the model therefore fixes a
coordinate system in which pretraining support can be mapped once, before any
task label is computed. The question posed to AL changes accordingly. Instead
of asking where the current model is uncertain, it asks which sampled task
environments have weak support in a fixed pretrained representation.
The support definition does not depend on the model being refined;
frame-level selection and geometric deduplication separately control which
of those environments are sent for labelling. This support diagnostic is distinct from a prediction-error estimate.

Here we introduce the Prior-guided Adaptive Refinement Scheme (PARS),
which connects a fixed pretraining-support reference with periodic
local data acquisition (Fig.~\ref{fig:workflow} and
Table~\ref{tab:comparison}).
PARS has two central methodological components.
(i)~\emph{Prior-guided acquisition.} A frozen teacher and
element-conditioned prototypes of the pretraining distribution
define support scores without initial task labels or an ensemble.
Both the representation and reference set remain fixed during
refinement, separating the assessment of pretraining support
from updates to the student potential.
(ii)~\emph{Periodic local reconstruction.} Selected environments
are converted into compact periodic cells through a search over
boundary placements constrained by local geometry and chemistry,
followed by relaxation of the movable surroundings using
the current student potential with the core fixed. DFT calculations on the resulting cells provide energy,
force and stress labels without requiring DFT calculations
on the full parent configurations.
Together, these components use pretrained knowledge to guide
where additional reference data are acquired and prepare local
environments for first-principles labelling.
The labels refine a student initialised from the foundation
checkpoint, which continues MD exploration while the acquisition
reference remains unchanged.

We examine the reference in high-pressure silicon, where support crossings
track coordination changes before any task label exists. We then test refined
potentials on a silicon benchmark and four property tests spanning
phase stability, surface hopping, vacancy clustering and interfacial exchange.
Finally, we distil a PARS-refined potential into a NEP without additional DFT training labels and
compute the TBR of a Si/Ge interface in a 362,880-atom cell. The predicted TBR agrees with
that obtained using a potential trained directly on DFT data. Together these tests separate
what the fixed reference detects, what refinement corrects, and how far the
resulting labels carry.

\section{Results}\label{sec:results}

\subsection{Separating the acquisition reference from the learner}\label{sec:framework}

\begin{figure}[p]
  \centering
  \includegraphics[width=\mainfigurewidth,height=0.70\textheight,keepaspectratio]{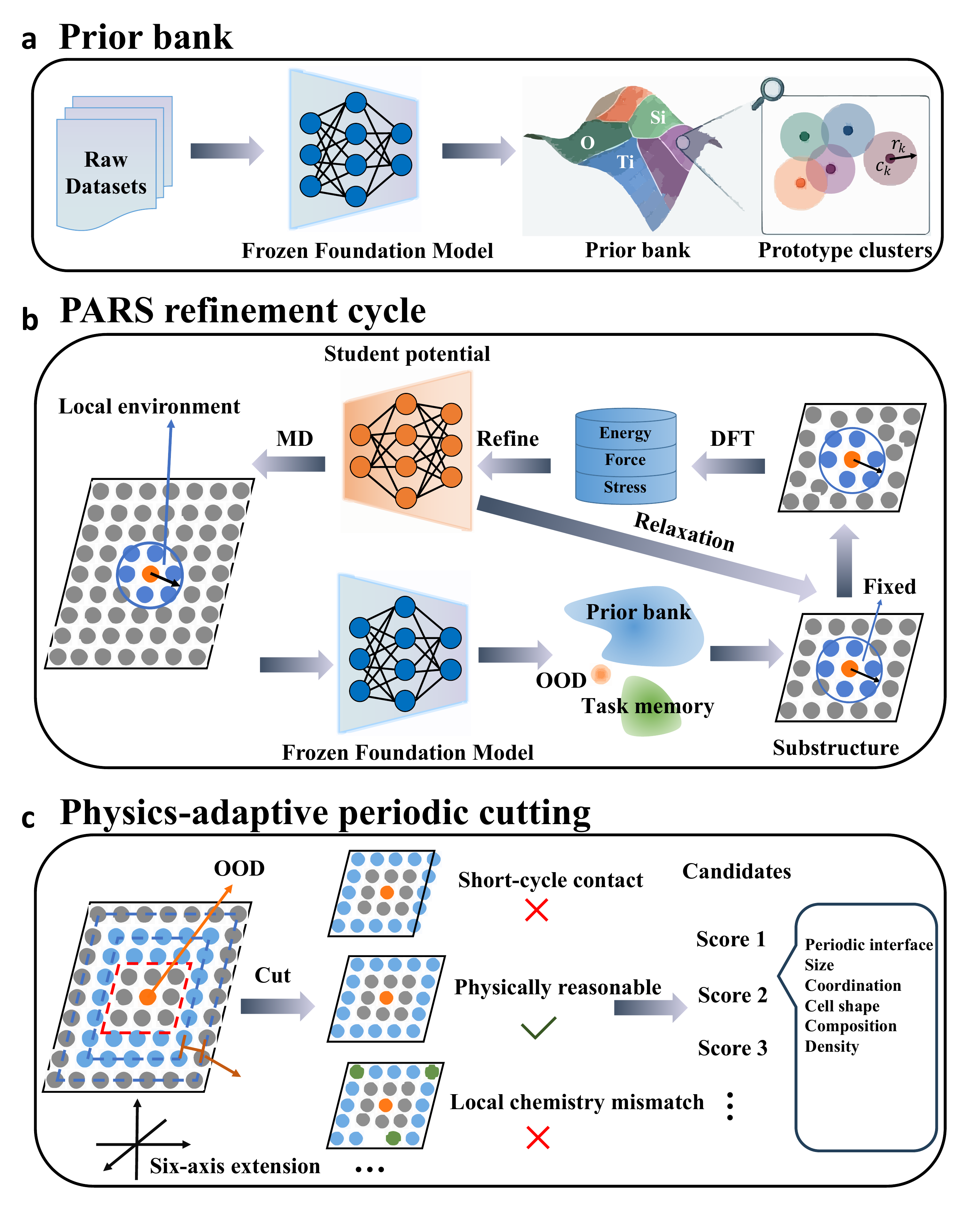}
  \caption{\textbf{PARS (Prior-guided Adaptive Refinement Scheme) framework.} (a) A frozen foundation model, the teacher, maps raw, element-labelled environments from the pretraining data to an element-conditioned prior bank of prototype centres and radii. (b) The refinement cycle. MD environments sampled by the student are scored against the frozen prior. At most one representative centre is retained per sampled frame; within-stream support ranking, interleaving and geometric deduplication control the batch. Acquisition records retain centre embeddings for provenance and diagnostics, not as a latent-distance gate. Selected centres are converted into periodic substructures, relaxed with a fixed core, labelled for energy, force and stress by first-principles calculations, and used to refine the student; the teacher and prior bank are never updated. (c) Physics-adaptive periodic cutting enumerates lower and upper face extensions along each non-preserved cell axis, rejects closures with short periodic contacts or a changed local chemistry around the centre, and ranks surviving closures by periodic-interface, size, coordination, cell-shape, composition and density terms. DFT denotes density-functional theory, and OOD denotes out-of-distribution.}
  \label{fig:workflow}
\end{figure}

PARS assigns two roles to two copies of one foundation model
(Fig.~\ref{fig:workflow}). The student is initialised from the pretrained
checkpoint, drives MD, repairs reconstructed cells and is refined. The
teacher is an unchanged copy of the same checkpoint and only embeds atomic
environments. For each element, prototypes of teacher embeddings drawn from
the pretraining data form a prior bank (Fig.~\ref{fig:workflow}a). An atom
whose embedding lies far from every same-element prototype, relative to the
prototype radii, has weak support. Its support score $s_{\mathrm P}$
(Eq.~\ref{eq:prior-score}) is compared with an element threshold that is
calibrated once on the starting configuration and then frozen. An atom
therefore becomes unsupported when the dynamics carry it beyond the support
level typical of its starting environments.

In each round PARS flags atoms whose support score exceeds the frozen
element threshold (Eq.~\ref{eq:acquisition-set}). Each sampled frame supplies
at most one representative centre, selected by its threshold-normalised
support score, and the frame representatives are ranked within each stream.
This construction separates two decisions (Table~\ref{tab:comparison}).
The support reference exists before the first label and needs one teacher
forward pass per sampled frame; no ensemble is trained. Accepted labels
change the student but neither this score nor its threshold, so labels cannot
relax the criterion that judges them. The event and geometry controls are
separate: a frame is reserved at most once within a cycle, and reconstructed
geometries matching previously committed labels are rejected before DFT.
Acquired embeddings are retained for provenance and optional distance
diagnostics, not as an exclusion gate in teacher space.

Different frames may therefore supply geometrically different cells with
similar embeddings. No packing-number bound on the total acquisitions is
claimed. A complete sampled cycle without an unsupported candidate is an
observation about those sampled frames, not a certificate that a region of
configuration space has been exhausted. Reaching a configured label, time or
generation limit is recorded as a budget stop, not as coverage convergence.

Selected atoms become labels through periodic reconstruction
(Fig.~\ref{fig:workflow}c). PARS does not label the parent cell. Instead, it
encloses a fixed core of radius $R$ around the centre in a periodic closure
chosen from six face extensions. Closures that break the core, alter the
centre's local chemistry or create short periodic contacts are rejected, and
the movable buffer is relaxed with the current student before DFT. The
resulting cells keep the parent core geometry and the declared boundary
conditions, and provide
stresses as well as energies and forces. In the Ti test, for example, 60--96-atom
cells were labelled from 768-site parent cells (Supplementary Table~S1).
Each MD stream, defined as one structure at one temperature, pressure and
phase, receives an equal initial share of the candidate budget. Unused shares
may be redistributed when explicitly enabled; otherwise they remain unused.
Candidates are interleaved across streams before reconstruction. Labels update
the student at configured training milestones, with a final update of any
untrained tail on normal termination. Each trajectory continues from its final
state using the active student, whether or not that cycle triggered an update.

\begin{table}[t]
\centering
\caption{\textbf{Selected acquisition strategies for interatomic
potential development and refinement.}
Entries describe the mechanisms used in the cited works;
implementations and extensions vary.
A fixed pretrained representation does not necessarily imply
a fixed acquisition reference: selection may also depend on
task labels, selected samples or an evolving predictive model.
The comparison concerns methodological design, not a
matched-budget assessment of accuracy or computational cost.}
\label{tab:comparison}

\footnotesize
\setlength{\tabcolsep}{2.5pt}
\renewcommand{\arraystretch}{1.15}

\begin{tabular}{@{}
>{\raggedright\arraybackslash}p{0.18\textwidth}
>{\raggedright\arraybackslash}p{0.29\textwidth}
>{\raggedright\arraybackslash}p{0.29\textwidth}
>{\raggedright\arraybackslash}p{0.16\textwidth}
@{}}
\toprule
\textbf{Approach}
&
\textbf{Acquisition signal}
&
\textbf{Reference used for acquisition}
&
\textbf{Local reconstruction}
\\
\midrule

D-optimality
(MTP)\cite{podryabinkin2017}
&
Extrapolation grade
&
Active set updated during learning
&
Not part of the cited acquisition criterion
\\

DP-GEN\cite{zhang2019dpgen,zhang2020dpgen}
&
Force disagreement across an ensemble
&
Models retrained on accumulated labels
&
Not part of the cited workflow
\\

FLARE\cite{vandermause2020flare}
&
Gaussian-process predictive variance
&
Posterior updated with reference observations
&
Not part of the cited workflow
\\

Latent-distance
estimators\cite{janet2019latent,tan2023uq}
&
Distance in a learned representation
&
Reference training data in model-dependent coordinates
&
Not part of the cited acquisition criteria
\\

DIRECT\cite{qi2024robust}
&
Representative sampling through dimensionality reduction,
clustering and stratification
&
Candidate structures represented by pretrained descriptors
&
Not part of the sampling method
\\

Pretrained-kernel
acquisition\cite{varga2026pretrained}
&
Activation and neural tangent kernels
&
Similarity information derived from a pretrained MACE model
&
Not part of the acquisition method
\\

DP-EVA\cite{liu2026dp}
&
Shallow-ensemble uncertainty combined with DIRECT sampling
&
Fine-tuned decoder ensembles and pretrained encoder descriptors
&
Not a focus of the cited framework
\\

Bayesian on-the-fly
fine-tuning\cite{rensmeyer2026fly}
&
Predictive uncertainty
&
Posterior updated from a prior informed by pretrained parameters
&
Not part of the cited workflow
\\

IDEAL\cite{kong2023ideal}
&
Force-field uncertainty
&
Uncertainty assessment with the current force field
&
Embedding of selected substructures
\\

\midrule

PARS (this work)
&
Distance to element-conditioned pretraining prototypes
&
Frozen teacher representation and fixed pretraining prototypes;
thresholds calibrated before refinement
&
Boundary search, fixed-core relaxation and validation
\\

\bottomrule
\end{tabular}
\end{table}

\clearpage
\clearpage

\subsection{Fixed support detects structural change before task labelling}\label{sec:si}

\begin{figure}[p]
  \centering
  \includegraphics[width=\mainfigurewidth]{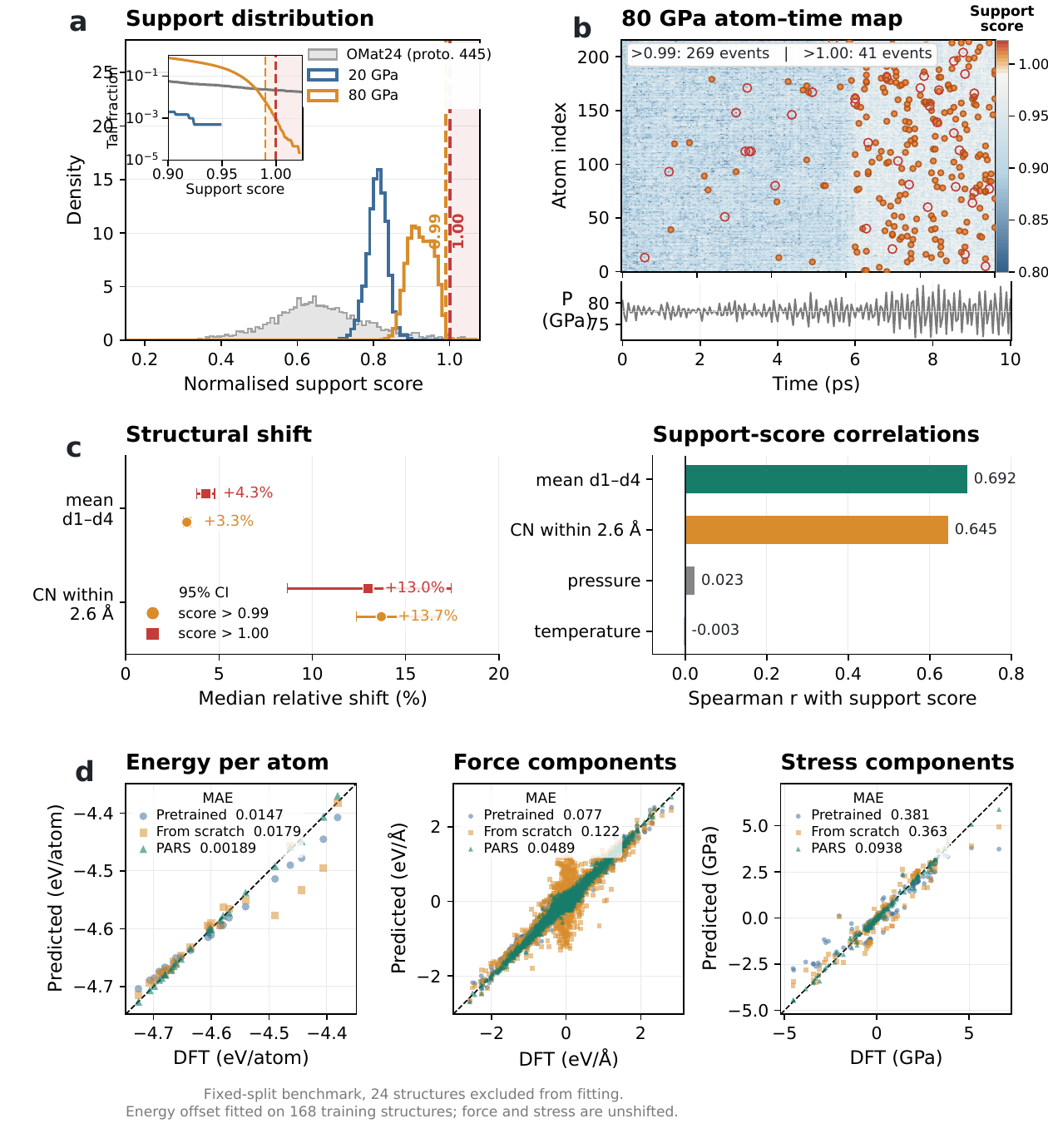}
  \caption{\textbf{High-pressure Si support diagnostics.}
  (a) Normalised support-score distributions for Si prototype 445 in the
  deduplicated Open Materials 2024 (OMat24) training set and 20/80\,GPa trajectories.
  The 0.99--1.00 interval is the monitoring band; scores above 1.00 cross
  the operational strict OOD boundary.
  (b) Atom--time map for the 80\,GPa trajectory (216 atoms, 201 frames).
  (c) Relative changes in coordination and neighbour geometry, with
  Spearman correlations to the support score. The d1--d4 metric is the
  mean distance to the first through fourth nearest neighbours.
  (d) DFT energy, force and stress parity for a refined
  checkpoint on 24 structures excluded from fitting. The energy metric uses a
  constant offset fitted only on the 168 training structures; forces and
  stresses are unshifted. MAE denotes mean absolute error, CN denotes
  coordination number, and OOD denotes out-of-distribution.}
  \label{fig:si}
\end{figure}

Because the reference precedes labelling, it can be examined on trajectories
that no refined model has seen. Figure~\ref{fig:si} uses a 512-prototype Si
support bank constructed from the deduplicated OMat24 training distribution. Panel (a)
compares support-score distributions for prototype 445 in training
environments and 20/80\,GPa isothermal--isobaric (NPT) trajectories. Among the 2,012
atom--frame evaluations assigned to this prototype at 20\,GPa, none crossed the
monitoring threshold ($s_{\mathrm P}>0.99$). At 80\,GPa, 269 monitoring events included 41
crossings of the strict out-of-distribution (OOD) boundary
($s_{\mathrm P}>1.00$) among 43,416 atom--frame evaluations. The first
monitoring crossing occurred near 0.35\,ps (Fig.~\ref{fig:si}a,b; Supplementary
Table~S3). Prototype radii were estimated from finite
cluster quantiles, so 2.2\% of the sampled training environments also scored
above one. The strict threshold therefore defines an operational support
boundary rather than absence from training, and the crossing frequency alone
does not separate this trajectory from training. The relevance of the score
rests on what the crossing atoms look like. These monitoring and strict
scores are $s_{\mathrm P}$ relative to the prototype radius; the calibrated
acquisition threshold of Supplementary Eq.~S5 need not coincide with them,
and only that threshold selects candidates.

Within the 80\,GPa trajectory, crossings followed local structure rather than
instantaneous pressure or temperature
(Fig.~\ref{fig:si}c). Relative to same-frame atoms with
$s_{\mathrm P}<0.99$, events showed a median 13.7\% increase in coordination
number (CN) within 2.6~\AA{} (95\% bootstrap confidence interval,
12.4--16.9\%). The mean first-to-fourth neighbour distance increased by
3.3\%. Spearman correlations with the support score were 0.65 for CN and
0.69 for this neighbour-distance metric, but only 0.02 for instantaneous
pressure and $-0.003$ for temperature. The fixed reference therefore
resolves which atoms enter weakly supported coordination environments, at
atomic and frame resolution, before any task label is computed. This is the
information an acquisition step needs; it is distinct from the predictive
error of the model, which is assessed separately below.

Two features of these crossings matter for acquisition. They were rare within
the trajectory: the
41 strict crossings correspond to 0.09\% of atom--frame evaluations and the
269 monitoring events to 0.6\%, so the atom-wise diagnostics flag a small
fraction of the sampled environments rather than entire configurations. They were also
transient and distributed, appearing in 100 of 201 frames and on 151 of 216
atoms rather than accumulating on a few sites (Fig.~\ref{fig:si}b). Repeated
excursions into similar local motifs motivate batch controls. In the current
workflow, the per-frame cap prevents a single sampled configuration from
providing multiple labels, and geometric deduplication rejects reconstructed
cells already committed. Similar embeddings in distinct geometries are not
automatically excluded; the crossing statistics do not measure label savings.

\clearpage
\clearpage

\subsection{Refinement corrects biases of the foundation potential}\label{sec:property}

We next evaluate previously generated refined potentials against DFT references
and physical observables. Each property model started from the same pretrained
checkpoint, with support-guided local labelling and periodic reconstruction,
using the historical, system-specific acquisition and training protocols
(Supplementary Table~S1 and Notes~3--6). These checkpoint evaluations assess the refined potentials,
rather than providing controlled comparisons of individual
acquisition and training choices. The property models used 756 (NaCl),
384 (NH$_3$/Pt), 192 (Ti) and 128 (LiI/Li$_2$S) DFT labels, a few hundred
small periodic cells per system.

\textbf{Silicon accuracy.} In a transfer
benchmark (Fig.~\ref{fig:si}d), 24 structures of a 192-structure Si dataset were
excluded from fitting and used as a fixed validation set. The refined checkpoint achieved energy, force
and stress mean absolute errors (MAEs) of 0.0019\,eV atom$^{-1}$,
0.049\,eV\,$\angstrom^{-1}$ and 0.094\,GPa, respectively. Corresponding
pretrained errors were 0.0147, 0.077 and 0.381 in the same units, and
from-scratch errors were 0.0179, 0.122 and 0.363. These values correspond to
reductions of 87\%, 36\% and 75\% relative to pretraining (Supplementary
Table~S4).

\begin{figure}[p]
  \centering
  \includegraphics[width=\mainfigurewidth]{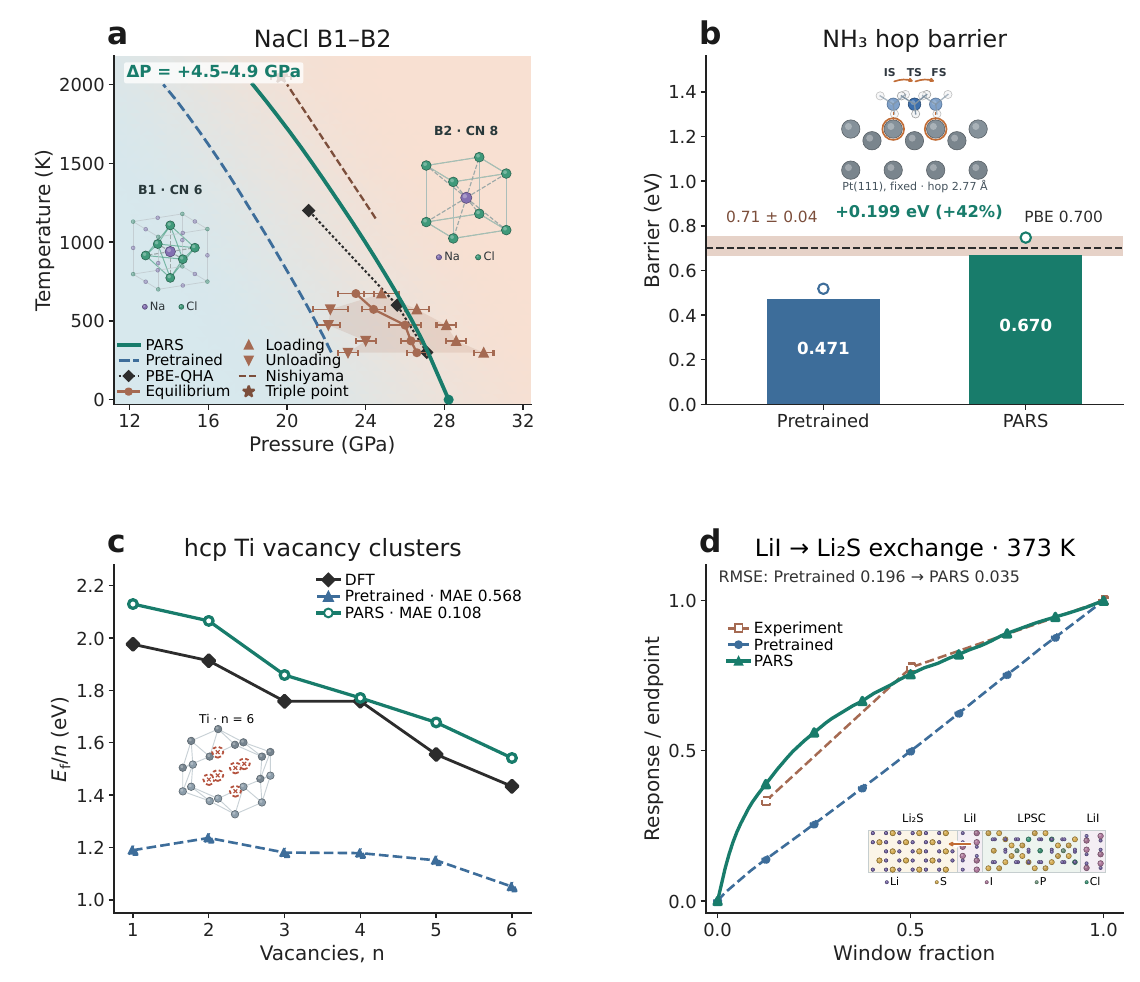}
  \caption{\textbf{Property benchmarks after task-specific refinement.}
  (a) NaCl coexistence boundary with PBE quasi-harmonic
  points~\cite{marcondes2018nacl}, Li--Jeanloz data~\cite{li1987nacl}
  and the Nishiyama X-ray fit and triple point~\cite{nishiyama2003nacl}.
  The background marks B1/B2 fields; the band spans loading--unloading
  hysteresis. The PARS boundary is anchored at 28.2\,GPa at 0\,K;
  error bars are statistical. Insets show Na-centred conventional
  cells: rock-salt B1 (six Cl neighbours, octahedral coordination) and
  CsCl-type B2 (eight Cl neighbours, cubic coordination). Purple and green
  mark Na and Cl; dashed radial lines highlight their nearest-neighbour
  contacts. CN denotes coordination number. Drawings are independently sized.
  (b) NH$_3$ hopping on a rigid four-layer, 4$\times$4 Pt(111) slab.
  Bars are ZPE-corrected barriers; open circles are raw barriers.
  The dashed line is the PBE reference (0.700\,eV); the band is
  $0.71\pm0.04$\,eV~\cite{borodin2021nh3}. In the inset, arrows connect
  adjacent on-top sites. IS, TS and FS denote initial, transition and final
  states. Translucent endpoint molecules are translated copies of the
  calculated TS pose, illustrating one molecule's hop; the upper Pt layers
  are shown and the slab remains fixed.
  (c) Formation energy per vacancy for $n=1$--6 hcp Ti clusters in a
  96-atom cell, on common DFT-relaxed geometries with each method's
  perfect-cell reference. Diamonds denote PBE/Ti\_pv, triangles the
  pretrained potential and open circles PARS; legend errors are MAEs
  in eV per vacancy. The local $n=6$ inset marks missing perfect-lattice
  sites with dashed rings; fine lines connect neighbouring Ti atoms.
  (d) LiI-to-Li$_2$S exchange at 373\,K, with an inset side view of the
  initial coated slab (LPSC denotes Li$_6$PS$_5$Cl). Squares denote EXSY intensities
  $I(t)/I(20\,\mathrm{s})$~\cite{liu2021lii}; circles and triangles
  denote pretrained and PARS MD probabilities $P(\tau)/P(100\,\mathrm{ps})$.
  The window fraction is $t/(20\,\mathrm{s})$ or $\tau/(100\,\mathrm{ps})$;
  physical times are not equated. Each MD curve uses one 1\,ns trajectory;
  Root-mean-square error (RMSE) values list pretrained then PARS.
  Structural arrows are guides.
  The 333/353\,K curves are in Supplementary Fig.~S1; lower-temperature
  shape errors are in Supplementary Table~S8.}
  \label{fig:property}
\end{figure}

Four property tests probe how refinement changes task-relevant properties
(Fig.~\ref{fig:property}): a Gibbs-free-energy crossing in NaCl, an NH$_3$
hopping barrier on Pt(111), Ti vacancy-cluster energetics and interfacial
exchange between LiI and Li$_2$S. They span the classes of configuration in
which universal potentials are known to be least reliable: compressed phases
with changed coordination, migration saddle points, defects, and interfaces
in which long-range electrostatics matter\cite{deng2025softening,zhang2022dplr,cheng2025les}.
The first three depend on small differences between large total energies and
are therefore sensitive to systematic bias.

\textbf{NaCl phase equilibrium.} Refinement shifted the estimated B1--B2
coexistence boundary towards higher pressure by 4.5--4.9\,GPa over
300--2,000\,K, while retaining its temperature dependence
(Fig.~\ref{fig:property}a). The competing phases are rock-salt B1 with
sixfold coordination and CsCl-type B2 with eightfold coordination. A
variable-cell enthalpy crossing anchored the PARS curve at 28.2\,GPa at
0\,K. Estimated coexistence pressures were 27.2, 25.9, 22.9 and
18.3\,GPa at 300, 600, 1,200 and 2,000\,K, respectively.
The Perdew--Burke--Ernzerhof (PBE) quasi-harmonic reference gives 27.1,
25.6 and 21.1\,GPa at the first three temperatures\cite{marcondes2018nacl}.
Against these values, the nominal three-point MAE decreased from 4.13 to
0.74\,GPa; the largest residual discrepancy was 1.83\,GPa at 1,200\,K.
The refined boundary also moved towards the loading--unloading constraints
of Li and Jeanloz\cite{li1987nacl} and the X-ray fit of Nishiyama
et al.\cite{nishiyama2003nacl}. Over 1,200--2,000\,K, the range of the X-ray fit, its average slope was
$-0.0058$\,GPa\,K$^{-1}$, compared with the experimental
$-0.0053$\,GPa\,K$^{-1}$. Both potentials were evaluated with the same
thermodynamic-integration protocol and cell, and neither contains an explicit
long-range term; statistical uncertainties and protocol details are given in
Supplementary Note~3 and Table~S5.

\textbf{NH$_3$/Pt(111) hopping.} Refinement raised the raw hopping barrier
from 0.518 to 0.747\,eV and the zero-point-energy (ZPE)-corrected barrier
from 0.471 to 0.670\,eV (Fig.~\ref{fig:property}b), an increase of
0.199\,eV (42\%). The discrepancy from the 4$\times$4 PBE reference of
0.700\,eV\cite{borodin2021nh3} decreased from 0.229 to 0.030\,eV, and the refined value
approached the lower edge of the experimental interval,
$0.71\pm0.04$\,eV\cite{borodin2021nh3}. The refined model used 384 labels
selected from ten MD streams of adsorbed, rotated and desorbing NH$_3$ between
300 and 1,500\,K. The benchmark used adjacent on-top
sites 2.77~\AA{} apart on a rigid four-layer, 4$\times$4 Pt(111) slab at
0.0625 monolayer coverage. Constrained relaxation and reaction-force
bisection located the saddle with Pt fixed; ZPE corrections included only
NH$_3$ modes. The shift is consistent with correcting the barrier softening
reported for universal potentials\cite{deng2025softening} (Supplementary
Note~4 and Table~S6).

\textbf{Ti vacancy-cluster energetics.} Refinement improved vacancy-cluster
formation energies and their size dependence in hexagonal close-packed (hcp)
Ti (Fig.~\ref{fig:property}c). The test contained one cluster for each size
from one to six vacancies in a 96-atom supercell. Both potentials were
evaluated on the same DFT-relaxed geometries, each using its own perfect-cell
energy reference. The refined model used 192 validated labels from 300\,K
trajectories seeded with single-plane vacancies; no test cell entered training.
The per-vacancy MAE decreased from 0.568 to 0.108\,eV, an 81\% reduction.
The pretrained potential underestimated formation energies by 0.38--0.79\,eV
per vacancy, whereas PARS overestimated them by 0.01--0.15\,eV.
From one to six vacancies, DFT formation energy per vacancy decreased from
1.976 to 1.433\,eV. The pretrained potential gave a shallower decrease,
from 1.189 to 1.050\,eV, while PARS decreased from 2.129 to 1.542\,eV.
Six-vacancy binding energies relative to isolated vacancies were 3.26\,eV
for DFT, 0.84\,eV for pretraining and 3.53\,eV for PARS.
The $n=2$, 5 and 6 clusters spanned two adjacent basal planes, a motif absent
from the refinement seeds, and retained comparable PARS errors
(Supplementary Note~5 and Table~S7).

\textbf{LiI-to-Li$_2$S interfacial exchange.} We simulated a 1026-atom
Li$_2$S/LiI/Li$_6$PS$_5$Cl/LiI slab for 1\,ns with each potential at five
temperatures from 293 to 373\,K. Phase assignments used the nearest anion
and required persistence for 2\,ps. The MD response was the probability that
Li initially in LiI occupied Li$_2$S after a lag, normalised by its value at
100\,ps. The experimental response was the $^6$Li two-dimensional exchange
spectroscopy (2D-EXSY) cross-peak intensity from Liu et al.\cite{liu2021lii},
normalised by its value at 20\,s. Scaling the two observation windows
independently compares curve shapes without equating physical times; the
common unit endpoint is imposed by normalisation. At 373\,K, refinement
changed the normalised response from an approximately linear rise to an
initially steeper curve that flattened towards the experimental shape, and
the root-mean-square error (RMSE) in normalised shape decreased from 0.196 to
0.035 (Fig.~\ref{fig:property}d). At 333 and 353\,K, shape RMSE decreased
from 0.135 to 0.031 and from 0.073 to 0.041, respectively (Supplementary
Fig.~S1). At 293 and 313\,K, it increased from approximately 0.066 to 0.109
and 0.120 (Supplementary Table~S8). Training used
103 structures from 313--373\,K, with all 25 structures at 293\,K held out,
and the refined model was trained on PBE-D3 labels together with a long-range
electrostatic term with learned atomic charges\cite{ji2025sog}, which the
pretrained model lacks. The improved exchange response at 333--373\,K is
therefore that of the refined model with its long-range term (Supplementary
Note~6).

Taken together, the first three tests share a pattern. In each, the
pretrained potential underestimated a quantity set by the energy of a
higher-energy or higher-coordinated configuration: the B1--B2 transition
pressure, the hopping barrier and the vacancy formation energy. This is
consistent with the systematic softening reported for universal
potentials\cite{deng2025softening}. Refinement removed most of the bias in
all three cases, leaving residuals of 0.74\,GPa (nominal MAE), 0.030\,eV and
0.108\,eV per vacancy, respectively, and in the Ti test it slightly
overcorrected. The LiI test differs in kind: it probes ion exchange at an
interface with a long-range contribution, and the improvement was confined to
333--373\,K. In every case the correction was obtained
with a few hundred small periodic labels.

\clearpage
\clearpage

\subsection{Periodic labels carry to a 362,880-atom interface}\label{sec:sige}

\begin{figure}[p]
  \centering
  \includegraphics[width=\mainfigurewidth]{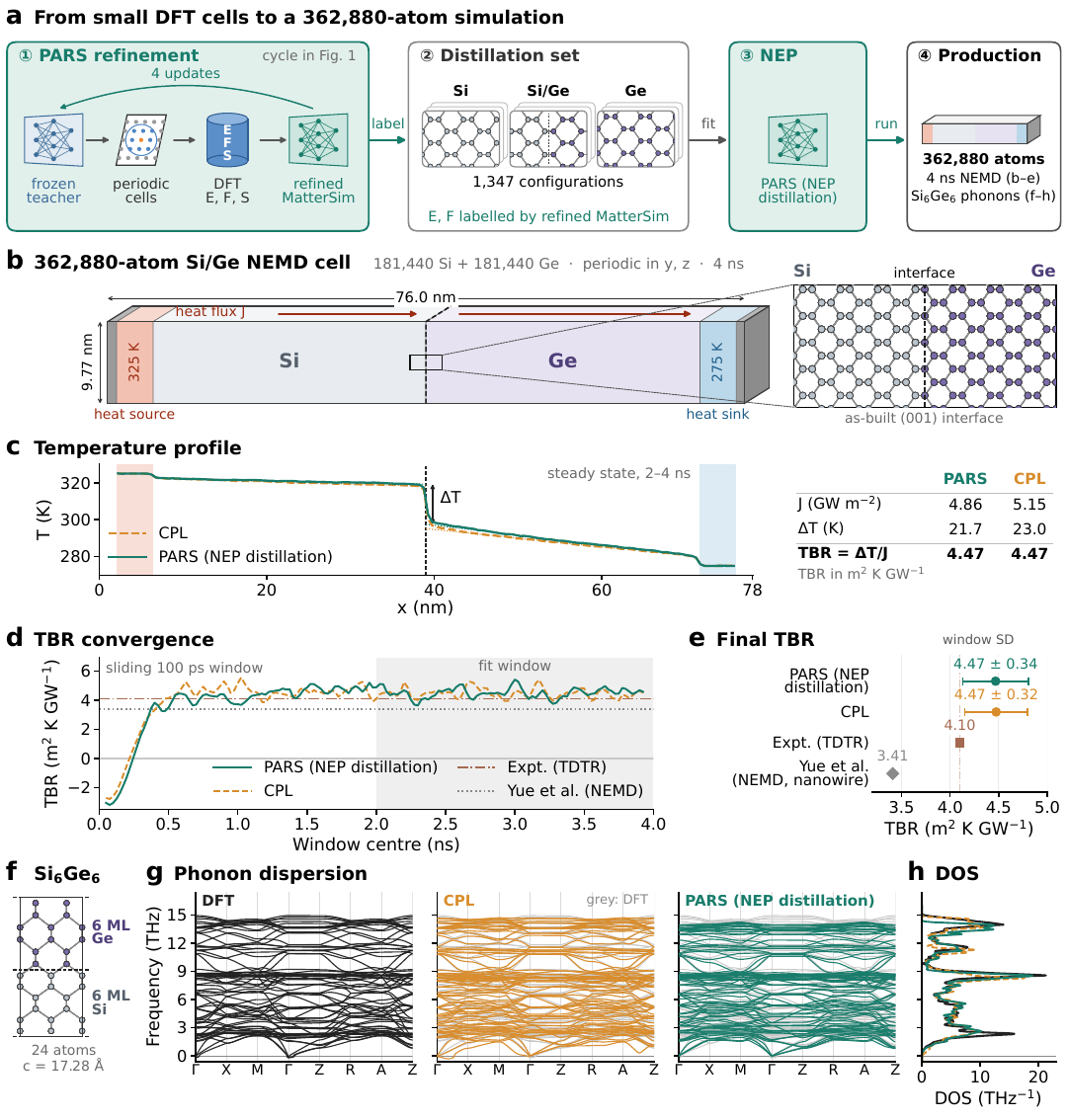}
  \caption{\textbf{Distilling a PARS-refined foundation potential for Si/Ge interfacial heat transport.}
  (a) DFT training labels enter only through PARS refinement (Fig.~\ref{fig:workflow}), in which fixed-core
  periodic cells receive DFT energy, force and stress labels. The refined MatterSim student then
  supplied energy and force labels for 1,347 configurations, to which a neuroevolution potential
  (NEP)\cite{fan2021nep}, PARS (NEP distillation), was fitted without further DFT labels (Methods). Card
  structures are representative.
  (b) NEMD cell, periodic along $y$ and $z$, with fixed ends (hatched) and 325/275\,K
  thermostats; the inset shows the as-built interface.
  (c) Temperature profiles averaged over 2--4\,ns for PARS (NEP distillation) and the reference CPL,
  a NEP trained directly on DFT data\cite{yue2025cpl}; dotted lines are the Si and Ge fits
  (6--24\,nm from the interface) that define $\Delta T$.
  (d) TBR in sliding 100\,ps windows; the final value is fitted over the shaded 2--4\,ns window.
  (e) Final TBR with experiment (4.10\,m$^2$\,K\,GW$^{-1}$; a second TDTR measurement gives 4.24)\cite{cheng2021tdtr} and nanowire NEMD\cite{yue2025cpl}. Error
  bars are the standard deviation of 100\,ps window estimates after 1.2\,ns and show
  window-to-window fluctuation, not the standard error of the mean.
  (f) DFT-relaxed Si$_6$Ge$_6$ superlattice. (g) Phonon dispersions on this geometry, plotted
  separately for DFT, CPL and PARS (NEP distillation); grey lines repeat DFT. (h) Total density
  of states ($8\times8\times8$ $q$ mesh, 0.08\,THz broadening). Force constants are from the
  24-atom cell.
  NEMD denotes non-equilibrium MD, TBR thermal boundary resistance and TDTR time-domain
  thermoreflectance.}
  \label{fig:sige}
\end{figure}

The tests above were evaluated at scales where the refined foundation model
itself can be run. Properties that need larger cells and longer times favour a faster potential.
The TBR of a Si/Ge interface is one: heat must cross the interface between
temperature gradients that develop over tens of nanometres and nanoseconds,
and the result depends on how the Si and Ge phonon spectra overlap. We
therefore asked whether labels from small periodic cells carry to such a
simulation once the refined potential is distilled into a fast architecture
(Fig.~\ref{fig:sige}a). The historical PARS campaign on Si/Ge
interfaces refined MatterSim through four model updates, during which DFT was evaluated
only on 800 reconstructed periodic cells of 10--128 atoms. The refined
MatterSim then labelled 1,347 Si/Ge configurations (194,624 atoms) with
energies and forces. A compact NEP, denoted
PARS (NEP distillation), was fitted to them without further DFT labels in
4.6\,h on eight GPUs, and its checkpoint was selected by a stability and DFT
diagnostic gate that involved no thermal-transport quantity (Methods). As a
reference we used CPL, a NEP trained directly on 739 DFT configurations
generated by empirical-potential annealing\cite{yue2025cpl}. Both potentials
were run through identical simulations, so differences between them reflect
the potentials rather than the protocol.

Non-equilibrium MD (NEMD) was performed on a 362,880-atom Si/Ge (001) cell,
76.0\,nm long, with heat source and
sink held at 325 and 275\,K (Fig.~\ref{fig:sige}b). Both potentials produced
linear temperature profiles in each material and a sharp drop at the
interface (Fig.~\ref{fig:sige}c). Over 2--4\,ns, PARS (NEP distillation)
carried a heat flux of 4.86\,GW\,m$^{-2}$ across a temperature jump of
21.7\,K, and CPL carried 5.15\,GW\,m$^{-2}$ across 23.0\,K. The TBR
evaluated in 100\,ps windows settled after a 0.5\,ns transient
(Fig.~\ref{fig:sige}d), and the final TBR was 4.47\,m$^2$\,K\,GW$^{-1}$ for
both potentials (Fig.~\ref{fig:sige}e). Across averaging windows their
difference ranged from $-0.1\%$ to $+2.1\%$, comparable to the standard error
of block averages, about 0.06\,m$^2$\,K\,GW$^{-1}$ for each potential
(Supplementary Table~S11).

The agreement is not inherited from identical bulk behaviour. The two
potentials differ by 6\% in heat flux, and their bulk temperature gradients
differ in opposite directions: relative to CPL, the distilled potential gives
a shallower gradient in Si and a steeper one in Ge (Supplementary
Table~S11). The apparent conductivities of the 33\,nm Si and Ge segments therefore
differ by about 20\% in each material, whereas the interfacial resistance
agrees to within statistical uncertainty. A plausible reading is that the TBR
is governed by vibrational states within a few nanometres of the interface,
which the interface configurations of both training sets sample
directly, whereas bulk conduction over tens of nanometres also depends on
long-wavelength phonons and their anharmonic scattering, which neither
potential was fitted to reproduce.
Both values lie 5--9\% above the time-domain thermoreflectance (TDTR)
measurements of 4.24 and 4.10\,m$^2$\,K\,GW$^{-1}$\cite{cheng2021tdtr}. Because the
offset is shared, it cannot be attributed to distillation; common
approximations, such as the PBE reference, classical phonon statistics or the
atomically sharp model interface, are more likely sources. The lower NEMD value of
3.41\,m$^2$\,K\,GW$^{-1}$ reported with CPL was obtained for a 45\,nm
nanowire\cite{yue2025cpl}.

Direct comparison with DFT, bypassing the refined MatterSim, supports the same
picture. Energies of Si/Ge interface cells deviated from DFT by $+0.2$
to $+0.4$\,meV\,atom$^{-1}$ (Supplementary Table~S10). On a 24-atom
Si$_6$Ge$_6$ superlattice, with one DFT-relaxed geometry for all methods
(Fig.~\ref{fig:sige}f--h), the mean absolute deviation of phonon frequencies
from DFT along the high-symmetry path was 0.23\,THz for PARS
(NEP distillation) and 0.18\,THz for CPL. Below 10\,THz, where the
acoustic and low-lying optical branches that carry most of the heat at 300\,K
reside, the deviations were 0.17 and 0.19\,THz, respectively. The larger
overall deviation of the distilled potential comes from optical branches above
10\,THz, which it places low by 0.38\,THz on average, with the highest
$\Gamma$ mode at 13.93 instead of 14.94\,THz; it has no imaginary modes away
from $\Gamma$. The
densities of states of both potentials overlap that of DFT to the same
extent, 0.82 (Supplementary Table~S12). Interfacial modes near 12\,THz, which
contribute about 5\% of the conductance across epitaxial Si--Ge
interfaces at 300\,K\cite{cheng2021tdtr}, lie in the softened band. That the TBR
nevertheless agrees suggests that this softening has a limited effect at
300\,K; a spectral decomposition of the interfacial heat current would be
needed to confirm it.

DFT training labels thus enter the TBR calculation only through periodic cells
more than three orders of magnitude smaller than the production simulation. The
production architecture is chosen afterwards and, for the quantities tested
here, remained close to DFT or to the DFT-trained reference.

\clearpage
\clearpage

\section{Discussion}\label{sec:discussion}

PARS changes the reference of active learning rather than the learner. A
frozen foundation representation and prototypes of its pretraining
distribution define where sampled task environments are unsupported. Acquisition
records retain centre embeddings, and periodic reconstruction turns selected
atoms into validated labels with stress. The acquisition signal depends on the
frozen prior, not the evolving student or a memory-distance veto. Acquisition
can start without task labels and its score cannot be relaxed by the labels it
produces. Frame-level selection, geometric deduplication and explicit resource
budgets control the campaign; they do not provide a sampling-independent bound
on the number of distinct environments that could be labelled. The results trace this
design through three levels: the reference detected coordination changes in
high-pressure silicon before any label existed, refinement corrected
systematic energetic biases of the foundation model, and a potential
distilled from the refined model reproduced an interfacial transport property
at a scale that DFT cannot reach.

The conventional AL criteria outlined above use an active
training set, an evolving predictive model or both to guide
acquisition\cite{podryabinkin2017,zhang2019dpgen,
vandermause2020flare,janet2019latent,tan2023uq}.
Foundation-potential refinement also permits acquisition
based on pretrained knowledge, as demonstrated by
pretrained-descriptor sampling and uncertainty-aware
fine-tuning workflows\cite{qi2024robust,varga2026pretrained,
liu2026dp,rensmeyer2026fly}.
PARS uses the same functional form as latent-distance
estimators, a distance to reference embeddings, but fixes
both the representation and the reference set: a frozen
pretrained model defines the coordinates, and prototypes
of pretraining environments define support.
This construction separates the assessment of pretraining
support from updates to the student potential and does not
require initial task labels. Whereas IDEAL selects local substructures using force-field
uncertainty\cite{kong2023ideal}, PARS uses fixed pretraining
support to identify candidate environments. Its reconstruction
procedure combines boundary-placement search, fixed-core
relaxation and validation before DFT calculations provide
energy, force and stress labels. At fixed core
protection and output budgets, PARS has lower measured prepared-embedding
time and peak process memory than the tested IDEAL implementation
(Supplementary Note~8, Supplementary Fig.~S2). This isolates the embedding
step rather than the complete acquisition--repair workflow and does not
establish equal electronic-structure fidelity.

The Si/Ge result places PARS within a hierarchy of models. DFT labels small
periodic cells, the refined foundation model generalises them across
configurations, and a compact potential takes that description to production
scale. Distillation of foundation models into faster architectures has been
demonstrated with synthetic data\cite{gardner2025distill} and with Hessian
matching\cite{amin2025hessian}; in these approaches, the DFT data come from
existing datasets or a separate sampling step, and distillation itself does not
choose them. PARS supplies that
step with a fixed criterion and periodic labels, and the distilled potential
matched a potential trained directly on DFT for the quantity of interest.
Interfacial resistance agreed although bulk conduction did not, which indicates
that agreement on one transport property does not certify others. Scoring
candidate distillation frames in the same frozen coordinates would place the
distillation set under the same fixed support diagnostic as the DFT labels.

The support score measures representation support, not predictive error. In
silicon it resolved coordination changes that instantaneous pressure and
temperature did not, which is the role required for acquisition. A fixed reference has a
complementary blind spot. Environments that are well represented in
pretraining but poorly predicted, for example because the task uses a
different exchange--correlation treatment from the pretraining data, do not
cross the support boundary. Conversely, fixed support scores can remain high
after a model has learned an environment, and geometrically distinct frames
with similar embeddings may still be selected. Geometric deduplication does
not guarantee novelty in representation space or low prediction error.
Error-based estimators could be added as a separate trigger for supported but
poorly predicted environments. The threshold
is anchored to the starting configurations, so a poorly supported start raises
it, and it must be recalibrated when the chemistry or thermodynamic range of a
campaign changes.

The property benchmarks were chosen to probe distinct failure modes of the
foundation model, each with its own established protocol: thermodynamic
integration for phase coexistence, a constrained saddle-point search on a
fixed slab for surface hopping, single cells per cluster size for vacancy
energetics, normalised exchange responses for interfacial transport and one
sharp (001) interface for the thermal boundary resistance. Extending them to
larger cells, rough or intermixed interfaces and independent trajectories, and
comparing PARS with random, descriptor-based and ensemble selection at equal
label budgets, are natural next steps for quantifying label efficiency across
chemistries.

Separating a fixed acquisition reference from an evolving potential turns
foundation-potential refinement into a campaign with a stable scoring target
and explicit budget and stopping records, rather than a coverage certificate.
Because the resulting labels are attached to small periodic cells
rather than to one architecture, a refined foundation model can also serve as
a label source for faster potentials, which brings properties such as
interfacial heat transport within reach of simulations informed by DFT labels
on small cells. The same separation could support refinement of other
foundation models, multi-task campaigns that share one reference, and
distillation sets assessed in the same frozen coordinates as the labels they
derive from.

\section{Methods}\label{sec:design}

The workflow combines a frozen representation prior, local structure
acquisition, first-principles labelling, property calculations and, for
Si/Ge, distillation into a neuroevolution potential. This section describes the PARS framework.
The acquisition and training protocols used for the
individual benchmarks are specified in the Supplementary
Methods and system-specific Supplementary Notes.
The complete formalism, parameter tables,
system-specific benchmark protocols and the Si/Ge settings are given in
Supplementary Methods~S1--S6 and Supplementary Tables~S1, S2 and S9--S12.

\subsection{Frozen reference and support score}

The student MLIP, $E_{\theta}(\R,\Z,\mathbf h)$, represents
the potential energy of a periodic configuration with
positions $\R$, species $\Z$ and cell matrix $\mathbf h$;
forces and stresses are energy derivatives.
All refined potentials start from the same MatterSim-based
checkpoint\cite{yang2025mattersim}, pretrained on
approximately seven million atomic configurations retained
from OMat24\cite{barros2026omat24} after similarity-based
deduplication using smooth overlap of atomic positions
(SOAP) descriptors.
The student drives production MD and local relaxation.
The teacher is a frozen copy of this checkpoint and embeds
atomic environments with its final graph-convolution layer.
The prior bank stores, for each element, prototype centres
$\mathbf c_{e,k}$ and radii $r_{e,k}$ obtained by offline
clustering of teacher embeddings of environments sampled
from the same deduplicated pretraining dataset.
For an atom $i$ with embedding $\mathbf z_i$ and species
$e_i$, the normalised support score is
\begin{equation}
  s_{\mathrm P}(i)=\min_{k}
  \frac{\lVert\mathbf z_i-\mathbf c_{e_i,k}\rVert_2^2}
       {r_{e_i,k}^2+\varepsilon},
  \label{eq:prior-score}
\end{equation}
with a small positive $\varepsilon$. Larger $s_{\mathrm P}$ means weaker
representation support relative to the fixed prototype scales; it is not an
energy or force error. Element thresholds $\tau_{e}$ are derived once from
the scores of the starting configurations (Supplementary Eq.~S5) and then
fixed, and an atom is unsupported when $s_{\mathrm P}(i)>\tau_{e_i}$. The
teacher, prior bank, thresholds and embedding code are verified before every
round (Supplementary Methods~S1).

\subsection{Event-level acquisition and acquisition records}

The unsupported atom set in sampled cycle $t$ is
\begin{equation}
  \mathcal A_t=\{\,i:\ s_{\mathrm P}(i)>\tau_{e_i}\,\}.
  \label{eq:acquisition-set}
\end{equation}
Neither labels nor stored embeddings enter Eq.~\ref{eq:prior-score} or
$\tau_e$. Optional natural-neighbour grouping produces candidate medoids;
without grouping, candidates are individual unsupported atoms. For each frame,
only the candidate with the largest $s_{\mathrm P}(i)/\tau_{e_i}$ is retained.
A bounded shortlist of frame representatives is ranked by within-element
support percentile, with the normalised score and enumeration order breaking
ties. There is no cosine-novelty term or task-memory distance exclusion.
Each stream receives an equal initial candidate allocation; unused allocations
may be redistributed when enabled, and selected candidates are interleaved.
Frame-event reservation applies within one cycle. After reconstruction and
repair, geometries duplicating committed labels or earlier prepared cells are
rejected using a translation- and atom-order-invariant fingerprint.
Centre embeddings are stored with accepted labels for provenance and optional
distance diagnostics. These controls do not imply a packing-number acquisition
bound (Supplementary Methods~S1).

\subsection{Periodic reconstruction and labelling}

Around each selected centre, a fixed core of radius $R$ and a movable buffer
are enclosed in a periodic closure chosen from six face extensions of the
parent cell. Closures with incomplete cores, a changed element-labelled
environment of the centre, or short periodic contacts are rejected; the
remaining closures are ranked by a physical score (Supplementary Methods~S2).
The student relaxes the buffer with the core and cell fixed, and the
structure is retained only if all force, stress, volume, distance and
displacement checks pass. Energies, forces and stresses are computed with a
consistent plane-wave projector augmented-wave setup in
VASP\cite{kresse1996,kresse1999} and validated before use (Supplementary
Methods~S3). The student is fine-tuned cumulatively on a Huber objective over
energies, forces and stresses at training milestones and is activated only
after validation; each MD stream then continues from its final coordinates,
cell and velocities with the active student. Existing validation members
remain excluded from fitting, while newly added labels may enlarge the
validation pool by a deterministic split. A campaign records sampled candidate
absence, geometric duplication and resource budgets separately; reconstruction
or labelling failures are not coverage convergence. On normal termination,
enabled training performs a final update if accepted labels remain unused by
the active model (Supplementary Methods~S3).

\subsection{Property calculations}

NaCl coexistence pressures follow Frenkel--Ladd thermodynamic
integration\cite{frenkel1984}, with the B1--B2 boundary as the root of
$\Delta G(P,T)=0$. The NH$_3$/Pt(111) barrier is the energy difference between
the constrained stationary point and the initial state, with a harmonic ZPE
correction from NH$_3$ modes. The formation energy of an $n$-vacancy Ti cluster
is $E_{\mathrm f}(n)=E_{N-n}-\frac{N-n}{N}E_{N}$ for $N=96$, with each method
referenced to its own perfect-cell energy on DFT-relaxed geometries. For the
LiI/Li$_2$S exchange, the MD endpoint probability and the 2D-EXSY intensity
are each normalised by their window endpoint and compared through a
shape RMSE. Protocols and settings are given in Supplementary Methods~S4 and
Table~S1.

\subsection{Distillation and Si/Ge heat transport}\label{sec:sige-methods}

For Si/Ge, the MatterSim potential was refined over four
model updates using 800 fixed-core periodic cells containing
10--128 atoms, labelled with PBE\cite{perdew1996pbe}. The refined MatterSim labelled 1,347
configurations with energies and forces, on which a NEP4
model\cite{fan2021nep,song2024nep4} with a Ziegler--Biersack--Littmark
short-range term\cite{ziegler1985zbl} was trained with
GPUMD\cite{fan2022gpumd}; the virial was not fitted. The checkpoint was
selected by the NVE stability of a Si/Ge bilayer and by DFT
diagnostics, not by training length or transport properties. NEMD used a
362,880-atom cell periodic along $y$ and $z$ with fixed ends and Langevin
source and sink at 325 and 275\,K; after 100\,ps of NVT
equilibration\cite{bussi2007}, 4\,ns were simulated with a 2\,fs time step.
The heat flux $J$ and the temperature profile were obtained from linear fits
and averages over 2--4\,ns; $\Delta T$ is the gap at the interface between
linear fits to the Si and Ge profiles 6--24\,nm from it, and
TBR $=\Delta T/J$. Phonons of the 24-atom Si$_6$Ge$_6$ superlattice used
finite-displacement force constants from phonopy\cite{togo2023phonopy} for
DFT and from GPUMD for the NEPs, on the same cell. The CPL potential was taken
from the authors' public repository\cite{yue2025cpl} and run with the same
cell and protocol. Full settings are given in Supplementary Methods~S6.

The implementation combines MatterSim, LAMMPS, ASE\cite{ase2017}, VASP, GPUMD and phonopy;
software and provenance are summarised in Supplementary Methods~S5 and
Table~S2.

\backmatter

\bmhead{Author contributions}
W.C., H.H. and J.L. conceived the study and jointly
directed the project.
Y.W., J.M., Y.Z., J.H.L., W.C., H.H. and J.L.
implemented the methods and workflows and performed
the computational tests and calculations.
Q.C., J.Y., W.Y., Y.C.Z. and C.Y. contributed to
low-level infrastructure development and computational
kernel optimisation.
All authors participated in discussions and contributed
to revising the manuscript.
J.L. led the research.

\bmhead{Competing interests}

The authors declare no competing interests.

\bibliography{fieldmat_pars_references}

\end{document}


\title[Supplementary Information]{Supplementary Information for ``Prior-guided adaptive local refinement of foundation interatomic potentials''}

\author[1]{\fnm{Yunlong} \sur{Wang}}\equalcont{These authors contributed equally to this work.}
\author[2]{\fnm{Jianzhen} \sur{Ma}}\equalcont{These authors contributed equally to this work.}
\author[1]{\fnm{Yusu} \sur{Zhang}}
\author[1]{\fnm{Jiahang} \sur{Li}}
\author[1]{\fnm{Qinqiu} \sur{Cheng}}
\author[1]{\fnm{Junjie} \sur{Yuan}}
\author[1]{\fnm{Wenrui} \sur{Yan}}
\author[3]{\fnm{Yucheng} \sur{Zhang}}
\author[3]{\fnm{Chen} \sur{Yu}}
\author*[2]{\fnm{Weibin} \sur{Chu}}\email{wbchu@fudan.edu.cn}
\author*[1]{\fnm{Hongxia} \sur{Hao}}\email{haohongxia@pjlab.org.cn}
\author*[1]{\fnm{Jielan} \sur{Li}}\email{lijielan@pjlab.org.cn}
\affil[1]{\orgname{Shanghai Artificial Intelligence Laboratory}, \city{Shanghai}, \postcode{200232}, \country{P. R. China}}
\affil[2]{\orgname{Key Laboratory for Computational Physical Sciences (MOE), State Key Laboratory of Surface Physics, Institute of Computational Physical Sciences and Department of Physics, Fudan University}, \postcode{200433}, \city{Shanghai}, \country{P. R. China}}
\affil[3]{\orgname{Hangzhou Research Center, Huawei Technologies Co., Ltd.}, \postcode{310052}, \city{Hangzhou}, \country{P. R. China}}

\abstract{This Supplementary Information describes the
PARS framework and the computational protocols used to
evaluate the reported models. Supplementary Methods
detail representation-guided acquisition, periodic
reconstruction, first-principles labelling, model refinement
and property calculations. Supplementary Notes provide
numerical results and clarify the scope and limitations
of the analyses presented in the main text.
}

\maketitle

\section{Supplementary Methods S1: Representation-guided acquisition}\label{supsec:acquisition}
\textbf{Pretrained model and reference dataset.}
The teacher and initial student share a MatterSim-based
checkpoint\cite{yang2025mattersim} pretrained on
approximately seven million atomic configurations.
This pretraining dataset was obtained by deduplicating
OMat24\cite{barros2026omat24} using structural similarity
assessed with smooth overlap of atomic positions (SOAP)
descriptors.

The prior bank is constructed from teacher embeddings
of atomic environments sampled from this same
deduplicated dataset. Accordingly, pretraining support
in this work refers to support relative to the retained
pretraining environments in the frozen model representation.
SOAP descriptors are used for dataset deduplication;
the PARS support score is computed from teacher embeddings,
not from SOAP descriptor distances.

\textbf{Environment embedding.} A periodic configuration is described by
atomic positions $\R=\{\mathbf r_i\}_{i=1}^{N}$, species
$\Z=\{Z_i\}_{i=1}^{N}$ and cell matrix $\mathbf h$. The student MLIP
represents a potential energy
\begin{equation}
  E_{\theta}=E_{\theta}(\R,\Z,\mathbf h),\qquad
  \mathbf F_i=-\nabla_{\mathbf r_i}E_{\theta},
  \label{eq:s-energy}
\end{equation}
with trainable parameters $\theta$; the stress tensor is obtained from the
cell derivative and is reported as six independent Voigt components using the
positive-tension convention. The student performs production MD and local
relaxation. A frozen teacher with parameters $\theta_{\mathrm T}$ supplies the
representation and is never updated during refinement. For atom $i$, let
$\rho_i$ denote the local periodic environment and $f_{\theta_{\mathrm T}}$
the final graph-convolution representation of the frozen teacher. The
environment embedding is
\begin{equation}
  \mathbf z_i=f_{\theta_{\mathrm T}}(\rho_i)\in\mathbb R^{d}.
  \label{eq:s-embedding}
\end{equation}
The MatterSim\cite{yang2025mattersim} extractor is environment-aware: embeddings are obtained from
the final graph layer with a 5~\AA{} graph cutoff and a 4~\AA{} three-body
cutoff. No student gradient is used to define the support score.

\textbf{Prior bank.} The prior bank is organised per chemical element,
\begin{equation}
  \mathcal B_{\mathrm{prior}}=\{\mathcal B_e\}_{e\in\mathcal Z},\qquad
  \mathcal B_e=\{(\mathbf c_{e,k},r_{e,k})\}_{k=1}^{K_e},
  \label{eq:s-bank}
\end{equation}
where $\mathbf c_{e,k}$ is a prototype centre and
$r_{e,k}\ge 0$ is its scale. The offline builder clusters
a reservoir of teacher embeddings of atomic environments
sampled from the deduplicated OMat24 pretraining dataset,
separately for each central element. Element conditioning restricts comparisons to environments
with the same central species, while retaining sensitivity
to local geometry and neighbouring chemical composition. Prototype radii are set from finite
cluster quantiles, which is why a small fraction of training environments can
score above one (Supplementary Note~2). For the Fig.~2 analysis the bank
holds $K=512$ Si centres (k-means\cite{lloyd1982kmeans}, seed 7); the production configuration uses
a canonical bank of 256 centres per element (seed 42). The normalised support
score for a query atom $i$ of species $e_i$ is
\begin{equation}
  s_{\mathrm P}(i)=\min_{k}
  \frac{\lVert\mathbf z_i-\mathbf c_{e_i,k}\rVert_2^2}
       {r_{e_i,k}^2+\varepsilon},
  \label{eq:s-prior-score}
\end{equation}
with positive numerical stabiliser $\varepsilon$. Larger $s_{\mathrm P}$
means weaker representation support relative to the fixed prototype scales.

\textbf{Thresholds.} For a finite score set $S_e$ collected during
calibration, the threshold statistic is
\begin{equation}
  b(S_e)=\alpha\,\operatorname{perc}_{p}(S_e)
       +(1-\alpha)\bigl(\operatorname{mean}(S_e)+\kappa\,\operatorname{std}(S_e)\bigr),
  \label{eq:s-threshold}
\end{equation}
and the prior threshold is $\tau_{\mathrm P,e}=a_{\mathrm P}b(S_e)$, where
$p$ is the percentile level, $\alpha$ blends the percentile and
mean--standard-deviation terms, $\kappa$ is the standard-deviation multiplier
and $a_{\mathrm P}$ is the prior scale. Only the prior is calibrated; neither
labels nor acquisition records re-tune the support boundary. By default
$S_e$ contains the scores of all element-$e$ atoms in the
starting configurations of all MD streams; a declared calibration set or
explicit thresholds can be supplied instead. The reference values are $p=95$, $\alpha=0.5$,
$\kappa=1$ and $a_{\mathrm P}=1$. An atom is therefore flagged when its support
is weaker than is typical of the starting environments. Thresholds are
computed once, recorded with the hashes of the calibration geometries, and then
frozen; recalibration is required when the chemistry or thermodynamic range
changes.

\textbf{Frozen prior mask.} The acquisition mask for atom $i$ of element
$e=e_i$ depends on the prior alone,
\begin{equation}
  m_i=\mathbf 1\{s_{\mathrm P}(i)>\tau_{\mathrm P,e}\},
  \label{eq:s-prior-gate}
\end{equation}
so an atom is unsupported whenever its support is weak relative to the
pretrained coverage. The mask is independent of the student revision and of
the accumulated labels.

\textbf{Acquisition records and optional memory diagnostic.} For each
committed label, the dataset records the teacher embedding $\mathbf m$ of the
acquired centre from its original parent frame. These records provide
provenance, not a veto on future acquisitions. Let $\mathcal M_e$ denote the
stored element-$e$ embeddings. The retained diagnostic class
\texttt{TaskMemory} can reconstruct, for each stored embedding, the radius
$r_{\mathbf m}=r_{e,k^*}$ of its nearest prior prototype,
$k^*=\operatorname{argmin}_k\lVert\mathbf m-\mathbf c_{e,k}\rVert$.
If explicitly evaluated, the corresponding diagnostic distance is
\begin{equation}
  s_{\mathrm M}(i)=\min_{\mathbf m\in\mathcal M_{e_i}}
  \frac{\lVert\mathbf z_i-\mathbf m\rVert_2^2}{r_{\mathbf m}^2+\varepsilon},
  \qquad s_{\mathrm M}(i)=\infty\ \text{if}\ \mathcal M_{e_i}=\varnothing .
  \label{eq:s-memory}
\end{equation}
The acquisition procedure described here does not evaluate or threshold this distance. In particular, it does not reject candidates with $s_{\mathrm M}\le1$, and
does not reserve exclusion balls in teacher space within a round. Only
accepted labels contribute stored acquisition records. Neither the records
nor Eq.~\ref{eq:s-memory} enters the support score, threshold or ranking.

\textbf{Unsupported atom set.} In sampled cycle $t$, atoms flagged by the
frozen support mask form
\begin{equation}
  \mathcal A_t=\{\,i:\ m_i=1\,\}.
  \label{eq:s-acquisition}
\end{equation}
Accepted labels can neither widen nor narrow the region that the prior
declares unsupported. The subset actually sent for reconstruction is
restricted by grouping, frame selection and the finite candidate budget.

\textbf{Acquisition limits.} Prototype radii are required to be positive, but the
acquisition procedure imposes no minimum embedding-space
separation between selected centres. Consequently, even
positive prior radii and bounded embeddings do not establish a packing-number
bound on the total number of its acquisitions. Finite campaign counts follow
from explicit label, MD-time, generation and internal cycle limits, not from
exhaustion of a latent-space memory. A no-candidate stop concerns the finite
frames sampled in that cycle. It is not a coverage or predictive-accuracy
certificate. Failed reconstruction and labelling are recorded separately
(Supplementary Methods~S3).

\textbf{Optional grouping and medoid.} The mask is atom-wise. When a dense
local event should not produce one label per atom, candidate atoms can be
grouped using
Atomic Simulation Environment (ASE)\cite{ase2017} natural-neighbour adjacency. For a
connected component $C$, the centre is the lexicographic medoid
\begin{equation}
  i^*=\underset{i\in C}{\operatorname{argmin}}\;
  \left(\sum_{j\in C}d_{ij}^{\mathrm{MIC}},
        \max_{j\in C}d_{ij}^{\mathrm{MIC}},i\right),
  \label{eq:s-medoid}
\end{equation}
where $d^{\mathrm{MIC}}$ is the minimum-image distance and the final index
makes ties deterministic. Components larger than the allowed group size are split
into bounded groups. Grouping is optional; without grouping,
each atom in $\mathcal A_t$ is an initial candidate.
In either case, the subsequent per-frame rule retains
at most one candidate from each sampled frame.

\textbf{Per-frame representative and per-stream objective.} Candidate scores
are normalised by the frozen element threshold,
$v_i=s_{\mathrm P}(i)/\tau_{e_i}$. Let $C_f$ be the initial candidates
(individual atoms or grouping medoids) in sampled frame $f$. Its
representative is
\begin{equation}
  i_f=\underset{i\in C_f}{\operatorname{argmax}}\ v_i,
  \qquad C_f\ne\varnothing .
  \label{eq:s-frame}
\end{equation}
Ties retain the first enumerated candidate. Each stream maintains a bounded
shortlist of its highest-scoring frame representatives. Within that
shortlist, normalised scores are converted into within-element percentiles
$p_i$, and candidates are ranked by
\begin{equation}
  J_i=p_i.
  \label{eq:s-pool}
\end{equation}
Ties are broken by descending $v_i$ and then enumeration order. No cosine
novelty term is evaluated. The legacy \texttt{ood\_weight} and
\texttt{novelty\_weight} configuration fields are retained as metadata but
are not used by the selection procedure described here.

For a cycle label budget $B$, bounded by the remaining label allowance and
the configured per-cycle cap, the candidate target is $T=\lceil1.6B\rceil$.
When the cap is null, the gap to the next training milestone sets $B$.
Each of $N_s$ streams receives an initial share $q=\lceil T/N_s\rceil$;
the default shortlist length is $\max(4q,32)$ unless explicitly overridden.
The selector reserves each (stream, frame) event once within the cycle.
If \texttt{redistribute\_unused} is enabled and the initial pass yields
fewer than $T$ candidates, remaining representatives compete for the unused
slots using the same support ranking. Otherwise those slots stay unused.
The selected lists are interleaved in stream order. Rounded stream shares
can initially exceed $T$; preparation processes at most $T$ candidates and
admits at most $B$ cells, including any recycled deferred cells.

After reconstruction and repair, a fingerprint of the cell, boundary
conditions, species and coordinates rejects geometries already committed or
already prepared in the batch. The fingerprint is invariant to atom order
and global translation, with cell and fractional coordinates rounded to six
decimal places. It does not establish that distinct geometries have distinct
embeddings, and frame reservations do not persist as exclusions across cycles.

\section{Supplementary Methods S2: Physics-adaptive periodic reconstruction}\label{supsec:cut}

\textbf{Closure enumeration.} The parent cell is repeated so that a sphere of
radius $R$ around the selected medoid and a maximum buffer extension are
represented. Six lower/upper face extensions
$\boldsymbol\delta=(\delta_1^-,\delta_1^+,\delta_2^-,\delta_2^+,\delta_3^-,
\delta_3^+)$ are generated from a finite distance grid. If $h_a$ is the
perpendicular height of parent axis $a$, a repeated atom with fractional
coordinate $s_{i,a}$ is retained when
\begin{equation}
  \frac{-R-\delta_a^-}{h_a}\le s_{i,a}<
  \frac{R+\delta_a^+}{h_a},\qquad a\in\{1,2,3\},
  \label{eq:s-closure}
\end{equation}
where $R$ is the fixed-core radius. Atoms within $R$ of the medoid form the
fixed core; all other retained atoms form the movable buffer. Axes that are
explicitly preserved, or are non-periodic in the parent, keep their original
extent and do not contribute a pair of face extensions; a two-dimensional
system therefore enumerates four extensions rather than six.

\textbf{Hard gates and physical ranking.} Candidate closures are rejected for
incomplete cores, an empty movable buffer, atom-count violations, degenerate
cells, a changed element-labelled environment of the centre (the species and
relative coordinates of all atoms within $R$ must match the parent), and
periodic short contacts. Contact limits are set for each element pair as the
parent minimum distance multiplied by a declared factor, bounded below by an
absolute floor, and the check includes contacts between an atom and its own
non-zero periodic images. Boundary coordination loss and the fixed-core fraction are
recorded for every candidate as diagnostics and can optionally be bounded per
system. Local composition and density enter the ranking below and can
additionally be bounded, because interfaces, defects and cells containing
vacuum should not be forced to reproduce the stoichiometry or density of the
bulk. For surviving candidates, each physical
term is min--max normalised over the valid set; a term whose spread is below
$10^{-3}$ of its largest magnitude is treated as constant and set to zero, so
that round-off cannot acquire full weight. Let $D_{\mathrm{pair}}$ denote the
periodic-pair fidelity, the fraction of neighbour pairs joined across a
periodic face of the closure for which the parent atom of the first member has
a same-element neighbour at the same separation vector, within
0.3\,\AA{} per Cartesian component (approximately the thermal displacement
scale). This term penalises artificial contacts created by the periodic
closure, whereas the coordination term penalises lost parent neighbours. Write
$D_N$ for target-size deviation and
$D_{\mathrm{coord}},D_{\mathrm{shape}},D_{\mathrm{comp}},D_{\mathrm{dens}}$
for the coordination, cell-shape, composition and density terms, the ranking
function is
\begin{equation}
  \begin{aligned}
  \Phi={}&2\bigl(1-\operatorname{norm}D_{\mathrm{pair}}\bigr)
     +2\operatorname{norm}D_N
     +\operatorname{norm}D_{\mathrm{coord}}\\
     &+\operatorname{norm}D_{\mathrm{shape}}
     +\operatorname{norm}D_{\mathrm{comp}}
     +\operatorname{norm}D_{\mathrm{dens}},
  \end{aligned}
  \label{eq:s-physical-score}
\end{equation}
and the lowest-$\Phi$ valid closure is selected once per centre. The relative
weights were fixed a priori for this study. Prepared-embedding time and peak
process memory, compared with our implementation of the
published IDEAL embedding procedure\cite{kong2023ideal},
are given in Supplementary Note~8 and Supplementary Fig.~S2.

\textbf{Constrained repair.} For cut $j$ with fixed atoms $C_j$ and movable
atoms $M_j$, the student calculator performs the constrained minimisation of
$E_\theta$ over the movable atoms subject to $\mathbf r_i=\mathbf
r_i^{(0)}$ for $i\in C_j$, using FIRE or BFGS. Reconstructions are retained
only if the optimiser converges without violating finite-force,
finite-stress, positive-volume, minimum-distance, core-displacement or
movable-atom displacement criteria. The cell is held fixed during the repair,
so the core keeps the parent coordinates and only the buffer is optimised.

\section{Supplementary Methods S3: First-principles labelling and student refinement}\label{supsec:label}

\textbf{Labelling contract.} First-principles labels use raw total energies
together with forces and stresses from a consistent plane-wave PAW--DFT setup
in VASP\cite{kresse1996,kresse1999}. A structure is admitted only when these quantities are finite, obey
the stated unit and positive-tension sign conventions, and use an energy
reference compatible with the other labels; the calculator energy and the
raw label energy must agree to $10^{-10}$\,eV. Incompatible energy references are rejected. Only successful validated
labels count towards training milestones and dataset targets.

\textbf{Training objective.} For a labelled structure $s$ with $N_s$ atoms,
the Huber objective\cite{huber1964robust} is
\begin{equation}
  \begin{aligned}
  \mathcal L={}&w_E\,\mathcal H_\delta\!\left(
      \frac{E_{\theta,s}-E_{s}^{\mathrm{DFT}}}{N_s}\right)
   +w_F\,\operatorname{mean}_{i,\alpha}\mathcal H_\delta\!\left(
      F_{\theta,s,i\alpha}-F_{s,i\alpha}^{\mathrm{DFT}}\right)\\
   &+w_\sigma\,\operatorname{mean}_{a,b}\mathcal H_\delta\!\left(
      \sigma_{\theta,s,ab}-\sigma_{s,ab}^{\mathrm{DFT}}\right),
  \end{aligned}
  \label{eq:s-huber}
\end{equation}
where $\mathcal H_\delta(x)$ is quadratic for $|x|\le\delta$ and linear
beyond. Stress enters as the symmetric tensor in training and is reported
with six Voigt components (eV\,$\angstrom^{-3}$, converted to GPa for
comparison). Hyperparameters and acquisition budgets are listed in
Table~\ref{tab:s-hyper}.

\textbf{Model updates.} Validated labels are retained cumulatively; each
successful DFT result is committed incrementally, while the next sampling
cycle still waits for the label stage, commit reconciliation and training
gate. A refined model is activated only after validation. A model generation
contains the sampling cycles between training updates. Training updates are triggered at specified cumulative
label counts. The initial update threshold and subsequent
growth ratio or fixed increment are configurable. The student is fine-tuned from the
previous checkpoint on cumulative training members; the selected checkpoint
minimises validation loss, rather than necessarily improving every reported
MAE. Each MD stream continues from the coordinates, cell and velocities of
its final state, with thermostat and barostat states reinitialised.

\textbf{Growing validation pool.} With the default
\texttt{stratified\_random} strategy, the first update selects validation
members within acquisition-stream strata using a seeded stable ranking.
Existing validation members remain excluded from training. On later updates,
each newly added sample is assigned deterministically by its stable identity
and seed, with probability set by \texttt{validation\_fraction}; samples
assigned to validation enlarge that pool and never return to training.
Thus membership decisions are persistent, but the validation set is not
frozen to its first-update contents. The grouped option similarly keeps
whole trajectory/cohort groups separated. These pools select checkpoints
and are not independent test sets. Optional external test data are
content-hash-pinned, checked for geometric overlap, and evaluated only after
checkpoint selection. The fixed 24-structure Si transfer benchmark reported
below is a separate historical evaluation, not this growing AL split.

\textbf{Stopping and final training.} A complete cycle with no unsupported
frame representative is recorded as \texttt{no\_new\_candidates}; a cycle
whose reconstructed geometries all duplicate committed labels is recorded
separately. Label, MD-time, generation and internal cycle budgets also have
distinct stopping reasons. A cycle without new labels or a triggered update
can terminate with \texttt{no\_new\_labels}, for example when all remaining
work was deferred; this is an operational stop, not coverage convergence.
If a bounded shortlist is exhausted without a usable candidate, or all
reconstructions or DFT tasks fail, the implementation raises an error rather
than interpreting failure as an empty unsupported set.

On normal termination, if training is enabled and the dataset contains labels
not yet used by the active model, a separate resumable \texttt{finalize/update}
stage trains on the cumulative split even below the next growth milestone.
Cancelled or failed campaigns do not automatically run this final stage.
No stopping reason certifies absence of unsupported environments outside the
finite sampled frames or supplies a latent-space packing bound.

\begin{table}[htbp]
\centering
\caption{Reference workflow settings and system-specific
calculation settings. Reference settings describe the
workflow in Supplementary Methods~S1--S3 and should not
be interpreted as the settings used for every benchmark.
System-specific protocols are given in the corresponding
rows and Supplementary Notes.}
\label{tab:s-hyper}
\begin{tabular}{@{}p{3.5cm}p{8.9cm}@{}}
\toprule
\textbf{Item} & \textbf{Setting} \\
\midrule
Pretraining dataset & Approximately seven million
OMat24 configurations retained after SOAP-based
similarity deduplication; the same dataset supplies
environments for prior-bank construction \\
Frozen representation & MatterSim final graph layer;
5/4\,\AA{} graph/three-body cutoffs for the stated model;
512 Si prototypes for Fig.~2 and 256 prototypes per
element in the reference bank \\
Support thresholds & Monitoring/strict scores 0.99/1.00 for Fig.~2; reference acquisition threshold $p=95$, $\alpha=0.5$, $\kappa=1$, $a_{\mathrm P}=1$ on the starting configurations \\
Reference acquisition settings & Prior-only mask; at most one representative per frame; support-percentile ranking; initial equal stream shares, candidate margin 1.6; singleton initial candidates; no latent-memory gate or cosine-novelty term \\
Reference training settings & Huber energy/force/stress weights $1/1/0.1$, $\delta=0.01$; at most 800 epochs, early-stop patience 5, mini-batch 64, learning rate $2\times10^{-4}$; persistent membership with a growing validation pool (fraction 0.1) \\
Reference update settings & 128 labels at the first update; 50\% growth; final training of the untrained tail on normal termination when training is enabled \\
NaCl refinement & 756 labels; isolated-atom-referenced energies; no long-range term \\
NaCl dynamics & 4,000 atoms; timestep 1\,fs; NPT equilibration/production 5/15\,ps; thermostat/barostat damping 0.1/1\,ps \\
NaCl free energy & Switching lengths 2/4/8\,ps; spring 8\,eV\,\AA$^{-2}$; one round and corrector; 2,000 bootstrap samples \\
NH$_3$/Pt(111) & Four-layer $4\times4$ slab, coverage 0.0625 monolayer; Pt fixed during hopping; molecular finite-difference displacement 0.02\,\AA \\
NH$_3$/Pt refinement & Ten MD streams, 300--1,500\,K; core radius 4\,\AA; 384 labels \\
NH$_3$/Pt DFT & PBE, 500\,eV; $k$-spacing 0.15\,\AA$^{-1}$; no dispersion correction \\
Ti refinement & 300\,K, 0\,GPa; 768-site cells; cuts of 60--96 atoms, fixed core radius 4\,\AA; 192 labels \\
Ti DFT & PBE, Ti\_pv, 400\,eV, $k$-spacing 0.3\,\AA$^{-1}$; smearing 0.1\,eV; test relaxation to 0.025\,eV\,\AA$^{-1}$ \\
Li exchange MD & 1026-atom slab; NPT 15\,ps, NVT equilibration 10\,ps, production 1\,ns; timestep 1\,fs; one trajectory per model and temperature \\
Li-interface DFT & PBE-D3, Li\_sv/I/S/P/Cl projectors, 520\,eV; $k$-spacing 0.2\,\AA$^{-1}$, smearing 0.05\,eV; 128 labels; long-range term included \\
Li refinement & NPT streams, 293--373\,K, 0\,GPa; core radius 3.5\,\AA; long-range term with learned charges \\
\bottomrule
\end{tabular}
\end{table}

\section{Supplementary Methods S4: Property calculations}\label{supsec:property}

\textbf{Thermodynamic integration.} The switching construction follows the
Frenkel--Ladd free-energy approach\cite{frenkel1984}. For coupling parameter $\lambda$, the
reversible free-energy difference is
\begin{equation}
  \Delta F=\int_{0}^{1}\left\langle\frac{\partial U_\lambda}
  {\partial\lambda}\right\rangle_{\lambda}\dd\lambda,
  \label{eq:s-ti}
\end{equation}
and the B1--B2 coexistence pressure is the root of $\Delta G(P,T)=0$. Model
parameters are held fixed throughout each complete forward/backward switching
round and its associated boundary correction; refinement is introduced only
between complete rounds. The local slope of the boundary is related to the
phase enthalpy and volume differences through the Clapeyron relation
$\dd P/\dd T=\Delta H/(T\,\Delta V)$.

\textbf{Zero-point-energy correction.} For the NH$_3$/Pt(111) calculation,
the raw hopping barrier is $\Delta E^\ddagger=E_{\mathrm{TS}}-E_{\mathrm{IS}}$,
where IS and TS denote the initial and transition states. The NH$_3$-only
harmonic correction is
\begin{equation}
  \Delta E^\ddagger_{\mathrm{ZPE}}=\Delta E^\ddagger+
  \frac{1}{2}\sum_{m\in\mathrm{TS}}\hbar\omega_m-
  \frac{1}{2}\sum_{m\in\mathrm{IS}}\hbar\omega_m,
  \label{eq:s-zpe}
\end{equation}
with the single imaginary transition-state mode excluded from the
real-frequency sum. Because the Pt slab is rigid during the hop, only
molecular modes enter the correction.

\textbf{Vacancy-cluster formation energy.} For a perfect periodic cell of
$N$ atoms with energy $E_{N}$ and a cell from which $n$ atoms have been
removed, with energy $E_{N-n}$, the formation energy is
\begin{equation}
  E_{\mathrm f}(n)=E_{N-n}-\frac{N-n}{N}\,E_{N},
  \label{eq:s-vac}
\end{equation}
and $E_{\mathrm f}(n)/n$ is reported per vacancy. Each method, whether DFT or
a potential, supplies both $E_{N}$ and $E_{N-n}$, so a constant per-atom
energy offset between methods cancels. The cells are relaxed at zero
pressure, so Eq.~\ref{eq:s-vac} is the zero-pressure formation energy. The
binding energy of the $n$-vacancy cluster relative to $n$ isolated vacancies
is $E_{\mathrm b}(n)=n\,E_{\mathrm f}(1)-E_{\mathrm f}(n)$, with positive
values denoting attraction.

\textbf{Exchange-curve shape.} Let $P(\tau)$ be the fraction of Li ions that
are assigned to LiI at the start of the production trajectory and to Li$_2$S
after a lag $\tau$. The molecular-dynamics response and the experimental
response are
\begin{equation}
  y_{\mathrm{MD}}(x)=\frac{P(100x\,\mathrm{ps})}{P(100\,\mathrm{ps})},\qquad
  y_{\mathrm{exp}}(x)=\frac{I(20x\,\mathrm{s})}{I(20\,\mathrm{s})},
  \label{eq:s-exchange}
\end{equation}
with the window fraction $x$ running from 0 to 1. The shape error is the
root-mean-square difference between $y_{\mathrm{MD}}$, linearly interpolated
in $x$, and $y_{\mathrm{exp}}$ at the three mixing times digitised from each
temperature in Liu et al.\cite{liu2021lii} Because both curves are divided by their own value
at $x=1$, that endpoint contributes a difference of zero. Equation
\ref{eq:s-exchange} compares shape only. It does not set 100\,ps equal to
20\,s, and it does not convert a probability into an NMR intensity.

\section{Supplementary Methods S5: Software and reproducibility}\label{supsec:software}

Table~\ref{tab:s-software} summarises the software used
in the calculations.

Supplementary Methods~S1--S3 describe the reference PARS
workflow, including event-level selection, cumulative
model refinement and validation. The model checkpoints
evaluated in Figs.~2--4 were obtained using system-specific
protocols. Their acquisition, energy-reference, data-split
and training settings are specified where available in
the corresponding tables and Supplementary Notes.
These benchmark evaluations should not be interpreted
as controlled comparisons of the reference acquisition
rule.

The numerical analyses below report support-score
statistics, coexistence pressures, hopping barriers,
vacancy energies, exchange responses and Si/Ge transport
and phonon results.

\begin{table}[htbp]
\centering
\caption{Software used for model refinement and property calculations.}
\label{tab:s-software}
\begin{tabular}{@{}p{3.5cm}p{8.9cm}@{}}
\toprule
\textbf{Software} & \textbf{Role} \\
\midrule
MatterSim-based potential\cite{yang2025mattersim}
& Model pretrained on the deduplicated OMat24 dataset;
frozen teacher representations and student refinement \\
PARS & Support-guided selection, periodic reconstruction and model refinement \\
LAMMPS\cite{thompson2022lammps} / ASE\cite{ase2017} & Molecular dynamics, geometry optimisation and finite-difference analysis \\
VASP 6.4.3\cite{kresse1996,kresse1999} & PAW--PBE\cite{perdew1996pbe} reference energies, forces and stresses \\
GPUMD 5.5\cite{fan2022gpumd} (\texttt{nep}) & NEP training, NEMD heat transport and NEP force constants (Si/Ge) \\
phonopy\cite{togo2023phonopy} & Finite-displacement DFT force constants and phonon dispersion (Si/Ge) \\
\bottomrule
\end{tabular}
\end{table}

\section{Supplementary Methods S6: Distillation and Si/Ge heat transport}\label{supsec:sige}

This section gives the full settings summarised in the main-text Methods. Data composition, NEP settings, checkpoint diagnostics, TBR sensitivity and phonon metrics are listed in Tables~\ref{tab:s-distil-data}--\ref{tab:s-phonon} (Supplementary Note~7).

\textbf{Si/Ge refinement campaign.}
The reported Si/Ge campaign used support-guided local
labelling with MD of Si/Ge interface cells over a range
of temperatures and pressures. Over four model updates,
800 fixed-core periodic cells containing 10--128 atoms
were labelled with VASP using the PBE
functional\cite{perdew1996pbe} and a 520\,eV plane-wave
cutoff. The resulting checkpoint is referred to as the
refined MatterSim.

\textbf{Distillation set.} The refined MatterSim supplied energies and forces;
no DFT label entered the NEP training set. The set contains 1,347
configurations with 194,624 atoms (Table~\ref{tab:s-distil-data}), drawn
mainly from MD of Si/Ge interface, bulk and mixed-composition cells; 173 of them, from targeted sampling of unstable
configurations, received a loss weight of 3. A validation set of 132
frames had no geometric overlap with the training set.

\textbf{NEP training.} A NEP4 model (Fan et al., 2021; Song et al., 2024) for Si and
Ge was trained with the \texttt{nep} executable of GPUMD~5.5\cite{fan2022gpumd}.
Radial and angular cutoffs were 5\,\AA, with $n_{\max}=12$ and 8, basis sizes
of 12 and $l_{\max}=4$ for three-body terms; four- and five-body terms were
disabled. The network had one hidden layer of 50 neurons, deliberately small
for speed. A Ziegler--Biersack--Littmark repulsion (Ziegler, Biersack and Littmark, 1985) with a
2\,\AA\ outer cutoff was added and switched off smoothly between 1 and
2\,\AA, keeping short contacts analytically repulsive. The energy weight was 1 and the virial weight 0, so stresses were
not fitted. Training used a population of 48 and batches of 100 for 200,000
generations, which took 16,460\,s on eight GPUs; training-set root-mean-square
errors at the end of training were 0.48\,meV\,atom$^{-1}$ and
34\,meV\,\AA$^{-1}$.

\textbf{Checkpoint selection.} The checkpoint was selected by a gate rather
than by training length (Table~\ref{tab:s-distil-nep}). The gate combined NVE
MD of a Si/Ge bilayer, checked for temperature stability, energy drift and
four-fold coordination of Ge, with DFT diagnostics, and contained no thermal-transport quantity.

\textbf{NEMD.} Simulations used GPUMD\cite{fan2022gpumd}.
The cell contained 70 conventional cells of Si and 70 of Ge along $x$
(76.0\,nm in total; 181,440 atoms of each element) and $18\times18$ cells
transversely (9.77\,nm; lateral lattice parameter 5.43\,\AA\ for both materials), and was periodic along $y$ and $z$, with 1\,nm of
vacuum beyond each end. End layers (9,720 atoms) were fixed. Source and sink
groups of 20,736 atoms adjacent to them were coupled to Langevin thermostats
at 325 and 275\,K with a 0.1\,ps coupling time. After energy minimisation and
100\,ps of NVT equilibration at 300\,K (Bussi--Donadio--Parrinello
thermostat (Bussi, Donadio and Parrinello, 2007), 0.5\,fs time step), 4\,ns were simulated with a
2\,fs time step. Temperatures were accumulated in 152 bins of 5\,\AA\ and
written every 1\,ps. The heat flux $J$ is the mean of the source and sink
energy-exchange rates, obtained by linear fits over 2--4\,ns, divided by the
95.5\,nm$^2$ cross-section. $\Delta T$ is the gap at the nominal interface
between linear fits to the Si and Ge profiles 6--24\,nm from it, and
TBR $=\Delta T/J$. The windowed TBR applies the same estimator to 100\,ps
windows with a 25\,ps step. Block uncertainties use non-overlapping 100\,ps
blocks after 1.2\,ns. The 100\,ps equilibration of 362,880 atoms ran at
$9.9\times10^{6}$ atom steps\,s$^{-1}$ on eight GPUs. The CPL potential file
was taken from the authors' public repository\cite{yue2025cpl} and run with
the same cell and protocol.

\textbf{Phonons.} The Si$_6$Ge$_6$ (001) superlattice (24 atoms,
$a=5.43$\,\AA, $c=17.276$\,\AA) was relaxed with PBE. DFT force constants were
obtained by finite displacements with phonopy\cite{togo2023phonopy} from 24
displaced structures (VASP, 520\,eV, $k$-point spacing 0.2\,\AA$^{-1}$,
electronic convergence $10^{-6}$\,eV). NEP force constants were computed with
GPUMD on the same 24-atom cell with 0.01\,\AA\ displacements. Dispersions
follow $\Gamma$--X--M--$\Gamma$--Z--R--A--Z. Densities of states use a
$\Gamma$-centred $8\times8\times8$ mesh with 0.08\,THz Gaussian broadening and
are normalised to 72 modes. Force constants of all three methods are computed
on the same 24-atom cell, so the comparison is like for like.

\section*{Supplementary Note 1: The PARS workflow (Fig.~1)}

The frozen teacher defines an element-conditioned coordinate system for
acquisition. Atoms are flagged when the pretrained prior provides
insufficient support (Eqs.~\ref{eq:s-prior-gate} and \ref{eq:s-acquisition}).
Frame-level support selection, stream interleaving and post-repair geometric
deduplication determine which candidates are labelled; no memory-distance
exclusion is applied. Selected centres are reconstructed as periodic fixed-core/movable-buffer
cells. Geometric screening and constrained relaxation precede DFT
labelling. Accepted energy, force and stress labels enter cumulative
student refinement; the teacher remains fixed. The construction and
selection rules are given in Eqs.~\ref{eq:s-bank}--\ref{eq:s-huber}.
Figure~1 describes the workflow; its accuracy and property consequences
are assessed separately in Figs.~2--4.

\section*{Supplementary Note 2: High-pressure silicon support diagnostics (Fig.~2)}

\textbf{Support analysis.} A 512-prototype Si bank is built from the
deduplicated OMat24\cite{barros2026omat24} training distribution using k-means (seed 7). The plotted
distributions and event map refer to prototype 445. The training sample
contains 2,602 scores, the 20\,GPa trajectory 2,012, and the 80\,GPa
trajectory 43,416 atom--frame evaluations (216 atoms, 201 frames).
No monitoring crossing occurs at 20\,GPa. At 80\,GPa, 269 events
cross 0.99 and 41 cross 1.00 (Table~\ref{tab:s-si-events}). Finite
cluster-quantile radii leave 56 of the 2,602 sampled training environments
(2.2\%) above one, so the threshold is an operational support boundary.

\textbf{Structural comparison.} Relative to same-frame supported atoms,
the monitoring events show increased coordination and mean first-to-fourth
neighbour distance. Their rank correlations with the support score are
0.6454 and 0.6923, compared with 0.0234 for pressure and $-0.0026$ for
temperature. The intervals in Table~\ref{tab:s-si-events} are bootstrap
95\% confidence intervals. These observations associate the score with
local geometry rather than instantaneous pressure or temperature.

\begin{table}[htbp]
\centering
\caption{Si support-event statistics for the 80\,GPa trajectory
(prototype 445).}
\label{tab:s-si-events}
\begin{tabular}{@{}p{7.7cm}p{4.7cm}@{}}
\toprule
\textbf{Quantity} & \textbf{Value} \\
\midrule
Trajectory & 216 atoms $\times$ 201 frames, 0--10\,ps \\
Pressure range (mean) & 71.881--84.264\,GPa (78.018\,GPa) \\
Temperature range (mean) & 1,244--1,779\,K (1,496\,K) \\
Monitoring events ($s_{\mathrm P}>0.99$) & 269 (100 frames, 151 atoms) \\
Strict events ($s_{\mathrm P}>1.00$) & 41 (37 frames, 37 atoms) \\
First monitoring crossing & $\approx$0.35\,ps \\
Maximum support score & 1.021710 \\
CN(2.6\,\AA) median relative change & $+13.701\%$ (95\% CI $12.363$--$16.894\%$) \\
Mean d1--d4 relative change & $+3.273\%$ (95\% CI $3.103$--$3.505\%$) \\
Strict subset: CN / d1--d4 change & $+12.999\%$ ($8.648$--$17.447\%$) / $+4.309\%$ ($3.798$--$4.764\%$) \\
Spearman(score, CN) & 0.6454 \\
Spearman(score, mean d1--d4) & 0.6923 \\
Spearman(score, pressure) & 0.0234 \\
Spearman(score, temperature) & $-0.0026$ \\
\bottomrule
\end{tabular}
\end{table}

\textbf{Transfer test.} A separate 192-structure dataset is divided into
168 training and 24 validation structures. A constant energy offset is fitted
on the training subset only; forces and stresses are unshifted.
Table~\ref{tab:s-si-mae} gives the errors on the fixed validation set. This
test assesses transfer accuracy and does not directly validate
the acquisition loop. The support events are candidate flags,
not completed DFT labels, and acquisition efficiency requires separate
comparisons at equal label budgets.

\begin{table}[htbp]
\centering
\caption{Errors for the Si transfer benchmark on the fixed validation set (24
structures excluded from fitting). Relative reductions are given against the pretrained and
from-scratch baselines.}
\label{tab:s-si-mae}
\begin{tabular}{@{}lccc@{}}
\toprule
\textbf{Model} & \textbf{Energy} & \textbf{Force} & \textbf{Stress} \\
\textbf{Units} & eV\,atom$^{-1}$ & eV\,\AA$^{-1}$ & GPa \\
\midrule
Pretrained & 0.014723 & 0.076958 & 0.380857 \\
From scratch & 0.017897 & 0.121930 & 0.362866 \\
PARS & 0.001889 & 0.048895 & 0.093826 \\
\midrule
Reduction vs pretrained & 87.17\% & 36.47\% & 75.36\% \\
Reduction vs from scratch & 89.44\% & 59.90\% & 74.14\% \\
\bottomrule
\end{tabular}
\end{table}

\section*{Supplementary Note 3: NaCl B1--B2 coexistence boundary (Fig.~3a)}

\textbf{Calculation.} The boundary uses 4,000-atom cells and one
forward/backward Frenkel--Ladd round followed by one corrector.
A variable-cell enthalpy calculation gives a 0\,K anchor of
$28.217963\pm0.003906$\,GPa. Settings are in Table~\ref{tab:s-hyper}.
The NaCl model uses an isolated-atom-referenced energy
convention, separate from the raw-energy label contract described above.

\textbf{Refinement.} The refined model was trained on 756 DFT labels; its
validation mean absolute errors are 1.4\,meV\,atom$^{-1}$ for energy,
0.020\,eV\,\AA$^{-1}$ for force and 0.045\,GPa for stress. Neither this
model nor the pretrained model contains an explicit long-range term.

\textbf{Comparison.} Table~\ref{tab:s-nacl} gives coexistence pressures
and references. The three-point MAE against PBE quasi-harmonic results
decreases from 4.129 to 0.739\,GPa. PARS shifts the pretrained curve
by 4.55--4.92\,GPa at the tabulated finite temperatures. Over 1,200--2,000\,K,
the range of the X-ray fit, the secant slopes are $-0.00584$ (PARS) and
$-0.00558$\,GPa\,K$^{-1}$ (pretrained), compared with the X-ray slope of
$-0.0053$\,GPa\,K$^{-1}$. PARS remains 1.31 and
1.75\,GPa below the X-ray fit at 1,200 and 2,000\,K.
The loading--unloading range represents experimental hysteresis.

\textbf{Uncertainties.} Reported errors describe pressure sampling with
2,000 bootstrap samples. The pretrained and refined boundaries were computed
with the same cell, switching schedule and corrector, so the shift between
them is a like-for-like comparison.

\begin{table}[htbp]
\centering
\caption{NaCl B1--B2 coexistence pressures (GPa). PBE quasi-harmonic values
from Marcondes et al.\cite{marcondes2018nacl}; the Nishiyama et al.\cite{nishiyama2003nacl} X-ray fit is
$P(\mathrm{GPa})=30.6-0.0053\,T(\mathrm{K})$.}
\label{tab:s-nacl}
\begin{tabular}{@{}lcccc@{}}
\toprule
\textbf{$T$ (K)} & \textbf{PARS} & \textbf{Pretrained} & \textbf{PBE-QHA} & \textbf{X-ray fit} \\
\midrule
0   & 28.218 (anchor) & --- & --- & --- \\
300 & 27.172 & 22.250 & 27.1 & --- \\
600 & 25.918 & 20.997 & 25.6 & --- \\
1,200 & 22.927 & 18.166 & 21.1 & 24.2 \\
2,000 & 18.254 & 13.705 & --- & 20.0 \\
\midrule
MAE vs PBE-QHA & 0.739 & 4.129 & --- & --- \\
Slope, 1,200--2,000\,K (GPa\,K$^{-1}$) & $-0.00584$ & $-0.00558$ & --- & $-0.0053$ \\
\bottomrule
\end{tabular}
\end{table}

\section*{Supplementary Note 4: NH$_3$/Pt(111) hopping barrier (Fig.~3b)}

\textbf{Calculation.} NH$_3$ hops between adjacent on-top sites
(2.771858\,\AA{} apart) on a four-layer $4\times4$ Pt(111) slab
at 0.0625 monolayer coverage. Pt is fixed during the hop. Constrained
relaxation and reaction-force bisection locate the saddle; molecular
finite differences of $\pm0.02$\,\AA{} give the ZPE correction
(Eq.~\ref{eq:s-zpe}), excluding the imaginary transition-state mode.

\textbf{Refinement.} The refined model was trained on 384 labels from ten MD
streams started from different adsorption sites and orientations of NH$_3$,
including desorbed configurations, at 300--1,500\,K. Labels used PBE
(500\,eV cutoff, no dispersion correction); settings are summarised in
Table~\ref{tab:s-hyper}. The 0.700\,eV PBE reference barrier is from Borodin et al.

\textbf{Comparison.} Refinement raises the raw and corrected barriers.
The corrected increase is 0.198768\,eV (42.19\%), reducing the
discrepancy from the 0.700\,eV PBE reference to 0.030\,eV
(Table~\ref{tab:s-nh3}). The refined barrier approaches the lower
edge of the $0.71\pm0.04$\,eV experimental interval. Both saddles
are near the hop midpoint and have one significant imaginary molecular
mode. This is a constrained rigid-slab benchmark; quantitative rates
require slab/coverage convergence and a fully relaxed pathway.

\begin{table}[htbp]
\centering
\caption{NH$_3$/Pt(111) on-top hopping benchmark (rigid four-layer
$4\times4$ slab, 0.0625 monolayer).}
\label{tab:s-nh3}
\begin{tabular}{@{}lcc@{}}
\toprule
\textbf{Quantity} & \textbf{Pretrained} & \textbf{PARS} \\
\midrule
Raw barrier (eV) & 0.518005 & 0.747070 \\
ZPE-corrected barrier (eV) & 0.471071 & 0.669839 \\
Saddle reaction coordinate & 0.501953 & 0.503906 \\
Residual reaction force (eV\,\AA$^{-1}$) & $1.21\times10^{-3}$ & $9.41\times10^{-6}$ \\
Imaginary frequency (meV) & $-14.30$ & $-14.92$ \\
\midrule
References & \multicolumn{2}{l}{PBE $4\times4$: 0.700\,eV; experiment: $0.71\pm0.04$\,eV} \\
Corrected increase & \multicolumn{2}{l}{$+0.198768$\,eV ($+42.19\%$), PARS vs pretrained} \\
\bottomrule
\end{tabular}
\end{table}

\section*{Supplementary Note 5: hcp Ti vacancy clusters (Fig.~3c)}

\textbf{Structures and reference.} The test contains one compact cluster
for each size of 1--6 vacancies in a 96-atom $4\times4\times3$ hcp
cell, selected through a preliminary screen of 869 configurations.
The final reference uses non-spin-polarised PBE with Ti\_pv, a 400\,eV
cutoff, 0.1\,eV smearing and a $3\times3\times2$ $k$-point mesh.
Atomic positions and cells are relaxed with DFT to
0.025\,eV\,\AA$^{-1}$. Both potentials are then evaluated on these
same geometries, each referenced to its own perfect-cell energy
(Eq.~\ref{eq:s-vac}).

\textbf{Refinement and comparison.} The refined model was trained on 192
labels from 300\,K, 0\,GPa NPT trajectories of vacancy-containing Ti cells,
with no test cell used in training. Its per-vacancy MAE
decreases from 0.568 to 0.108\,eV and its six-vacancy binding energy
approaches DFT (Table~\ref{tab:s-ti}). Test clusters with 2, 5 and 6
vacancies span two basal planes, unlike the refinement seeds.
The benchmark measures energetic predictions on common geometries.
DFT relaxation starts from refined-model minima, and finite-size,
additional $k$-point and functional convergence are untested; results
apply to this six-structure reference protocol.

\begin{table}[!htbp]
\centering
\caption{Formation energy per vacancy, $E_{\mathrm f}(n)/n$ (eV), for the
hcp Ti test clusters, and binding energy of the largest cluster. All values
are single points on the DFT-relaxed geometries.}
\label{tab:s-ti}
\begin{tabular}{@{}lccc@{}}
\toprule
\textbf{$n$} & \textbf{DFT (Ti\_pv)} & \textbf{Pretrained} & \textbf{PARS} \\
\midrule
1 & 1.976 & 1.189 & 2.129 \\
2 & 1.913 & 1.236 & 2.065 \\
3 & 1.758 & 1.180 & 1.858 \\
4 & 1.758 & 1.178 & 1.771 \\
5 & 1.555 & 1.151 & 1.677 \\
6 & 1.433 & 1.050 & 1.542 \\
\midrule
MAE vs DFT & --- & 0.568 & 0.108 \\
Mean signed error & --- & $-0.568$ & $+0.108$ \\
RMSE vs DFT & --- & 0.585 & 0.118 \\
$E_{\mathrm b}(6)$ (eV) & 3.26 & 0.84 & 3.53 \\
\bottomrule
\end{tabular}
\end{table}

\FloatBarrier
\section*{Supplementary Note 6: LiI/Li$_2$S lithium exchange (Fig.~3d and Supplementary Fig.~S1)}

\textbf{Simulation and response.} A periodic 1026-atom
Li$_2$S/LiI/Li$_6$PS$_5$Cl/LiI slab is simulated at 293, 313, 333,
353 and 373\,K. Each model has one 1\,ns trajectory per temperature,
with paired initial seeds. Li phase assignment uses the nearest non-Li
neighbour and a 2\,ps persistence requirement. Exchange responses are
normalised by their 100\,ps endpoint; EXSY intensities are normalised
by their 20\,s endpoint (Eq.~\ref{eq:s-exchange}). Experimental points
are manually digitised near 2.5, 10 and 20\,s from Liu et al., Fig.~4b;
shorter overlapping markers are omitted and no measurement error bars
are available. The shape RMSE uses these three points, including the
normalised endpoint whose difference is zero.

\textbf{Refinement and long-range term.} Candidates were sampled from NPT
trajectories of the coated slab at 293--373\,K and labelled with PBE\cite{perdew1996pbe}-D3\cite{grimme2010d3}
(Table~\ref{tab:s-hyper}); the plotted model used 128 labels, of which the 25 at 293\,K were held out. Its
long-range term predicts atomic charges with a small neural network and adds
a periodic electrostatic energy; it was trained together with the
short-range model during refinement, following the sum-of-Gaussians (SOG) construction\cite{ji2025sog}.

\begin{figure}[p]
\centering
\includegraphics[width=0.96\textwidth]{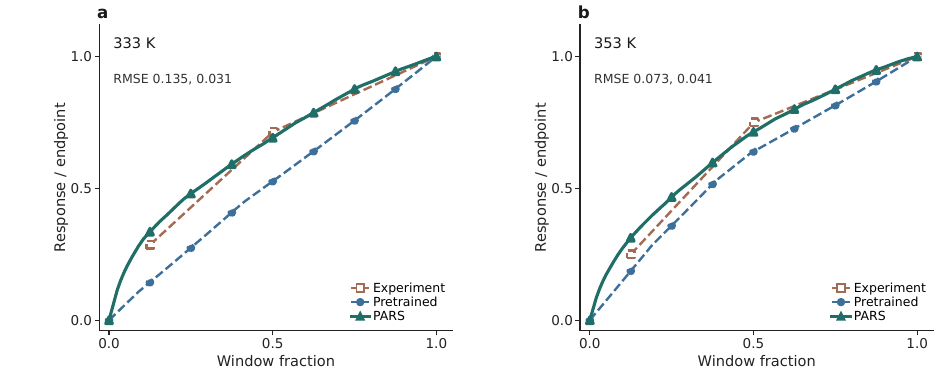}
\caption{\textbf{LiI-to-Li$_2$S exchange responses at additional temperatures.}
(a) 333\,K and (b) 353\,K. Squares denote experimental 2D-EXSY intensities
from Liu et al.; circles and triangles denote MD endpoint probabilities for
the pretrained potential and PARS. The window fraction is the MD lag
divided by 100\,ps or the experimental mixing time divided by 20\,s;
each response is independently normalised by its window endpoint
(Eq.~\ref{eq:s-exchange}). Physical times are not equated. Each MD curve
uses one 1\,ns trajectory; RMSE values list pretrained then PARS.
The 373\,K comparison is shown in main-text Fig.~3d.}
\label{fig:s-exchange}
\end{figure}

\textbf{Comparison and scope.} Table~\ref{tab:s-lii} reports all five
temperatures. PARS improves the shape at 373\,K (Fig.~3d), and at
333 and 353\,K (Supplementary Fig.~\ref{fig:s-exchange}), but worsens it
at 293 and 313\,K. Training used 103 structures from 313--373\,K; all 25 structures at
293\,K were held out. Refined runs include the long-range term described
above, which pretrained runs lack. The comparison concerns the normalised
exchange response over separately scaled time windows.

\begin{table}[htbp]
\centering
\caption{LiI$\to$Li$_2$S exchange at five temperatures. Shape errors use
Eq.~\ref{eq:s-exchange}. Figure~3d shows 373\,K; Supplementary
Fig.~\ref{fig:s-exchange} shows 333 and 353\,K.}
\label{tab:s-lii}
\begin{tabular}{@{}lcc@{}}
\toprule
\textbf{$T$ (K)} & \shortstack{\textbf{Pretrained}\\\textbf{shape RMSE}} &
\shortstack{\textbf{PARS}\\\textbf{shape RMSE}} \\
\midrule
293 & 0.066014 & 0.108598 \\
313 & 0.065991 & 0.120112 \\
333 & 0.135163 & 0.031464 \\
353 & 0.072976 & 0.041477 \\
373 & 0.195734 & 0.034941 \\
\bottomrule
\end{tabular}
\end{table}

\clearpage

\clearpage
\section*{Supplementary Note 7: Si/Ge distillation and heat transport (Fig.~4)}

\textbf{Distillation chain.} Settings are given in Supplementary
Methods~S6. The refined MatterSim labelled the distillation set with energies
and forces (Table~\ref{tab:s-distil-data}); the set contains no DFT labels
and no virials. The reference potential CPL is the NEP of Yue et al.\cite{yue2025cpl}
(Chin. Phys. Lett. \textbf{42}, 036301 (2025)), trained on 739 PBE
configurations from Stillinger--Weber annealing of bulk Si, bulk Ge and
Si/Ge heterostructures.

\textbf{Thermal boundary resistance.} Both potentials were run in the same
362,880-atom cell with the same protocol (main-text Methods). The estimator
of Fig.~4c--e uses linear fits of the thermostat energies and the mean
temperature profile over 2--4\,ns. Table~\ref{tab:s-tbr} repeats the
estimate for other windows and for block averages. The PARS--CPL difference
ranges from $-0.1\%$ to $+2.1\%$ and is comparable to the combined standard
error of the block means (0.07--0.08\,m$^2$\,K\,GW$^{-1}$). The two
potentials differ more in the bulk temperature gradients than in the
interfacial resistance. Fig.~4e error bars are the standard deviation of the
overlapping 100\,ps window estimates after 1.2\,ns (0.336 and
0.323\,m$^2$\,K\,GW$^{-1}$); they describe window-to-window fluctuation, not
the uncertainty of the mean. Before 0.25\,ns the windowed TBR is negative
for both potentials because the temperature profile is still building up.

\textbf{Phonons.} Table~\ref{tab:s-phonon} summarises the comparison in
Fig.~4g,h. Frequencies of CPL and PARS (NEP distillation) are interpolated
onto the DFT path coordinates and compared band by band after sorting.
The force constants of all three methods are computed on the same 24-atom
cell, so the comparison is like for like.

\begin{table}[htbp]
\centering
\caption{Distillation set for PARS (NEP distillation). System counts are
approximate.}
\label{tab:s-distil-data}
\begin{tabular}{@{}p{3.6cm}p{8.8cm}@{}}
\toprule
\textbf{Item} & \textbf{Value} \\
\midrule
Label source & Refined MatterSim (after the fourth update) \\
Labels & Energies and forces; no DFT labels; no virials \\
Training set & 1,347 frames, 194,624 atoms \\
Validation set & 132 frames, no geometric overlap with the training set \\
Frame weights & 1 for MD samples; 3 for 173 frames from targeted sampling of unstable configurations \\
\midrule
\multicolumn{2}{@{}l}{\textit{By system (approximate)}} \\
Si/Ge interfaces & $\approx$460 (200--400\,K; strained; intermixed) \\
Bulk Si and Ge & $\approx$246 \\
Si$_{1-x}$Ge$_x$ alloys & 280 ($x=0.25$, 0.5, 0.75) \\
Other & $\approx$361 \\
\bottomrule
\end{tabular}
\end{table}

\begin{table}[htbp]
\centering
\caption{NEP architecture, training and checkpoint selection. Diagnostics
refer to the selected checkpoint; DFT diagnostics compare the NEP directly
with PBE.}
\label{tab:s-distil-nep}
\begin{tabular}{@{}p{4.4cm}p{8.0cm}@{}}
\toprule
\textbf{Item} & \textbf{Value} \\
\midrule
Model & NEP4, species Si and Ge; ZBL outer cutoff 2.0\,\AA \\
Descriptor & Cutoffs 5/5\,\AA; $n_{\max}=12/8$; basis size 12/12; $l_{\max}=4/0/0$ \\
Network & One hidden layer, 50 neurons \\
Loss & $\lambda_e=1$, $\lambda_v=0$; frame weights 1 and 3 \\
Optimiser & SNES, population 48, batch 100, 200,000 generations \\
Training cost & GPUMD~5.5 \texttt{nep}, 8 GPUs, 16,460\,s (4.6\,h) \\
Training RMSE (end of run) & 0.48\,meV\,atom$^{-1}$; 34\,meV\,\AA$^{-1}$ \\
\midrule
Checkpoint selection & Gate on Si/Ge bilayer NVE stability and DFT diagnostics \\
\midrule
\multicolumn{2}{@{}l}{\textit{Selected checkpoint}} \\
Bulk Si & $-5.4154$\,eV\,atom$^{-1}$; max.\ residual force $6.5\times10^{-6}$\,eV\,\AA$^{-1}$ \\
Bulk Ge & $-4.4600$\,eV\,atom$^{-1}$; max.\ residual force $5.4\times10^{-6}$\,eV\,\AA$^{-1}$ \\
Interface energy vs DFT & $+0.2$ to $+0.4$\,meV\,atom$^{-1}$ \\
\bottomrule
\end{tabular}
\end{table}

\begin{table}[htbp]
\centering
\caption{Sensitivity of the TBR to the averaging window. $J$ in
GW\,m$^{-2}$, $\Delta T$ in K, TBR in m$^2$\,K\,GW$^{-1}$; $\nabla T$ is the
fitted bulk temperature gradient in K\,nm$^{-1}$. Block rows give mean
$\pm$ standard error. The difference is (PARS$-$CPL)/CPL. $^{a}$Apparent conductivity of each 33\,nm segment in this cell; size-limited, not a bulk value.}
\label{tab:s-tbr}
\footnotesize
\setlength{\tabcolsep}{3pt}
\begin{tabular}{@{}lcccccccc@{}}
\toprule
 & \multicolumn{3}{c}{\textbf{PARS (NEP distillation)}} & \multicolumn{3}{c}{\textbf{CPL}} & \\
\cmidrule(lr){2-4}\cmidrule(lr){5-7}
\textbf{Estimator} & $J$ & $\Delta T$ & TBR & $J$ & $\Delta T$ & TBR & \textbf{Diff.} \\
\midrule
Fit, 2--4\,ns (Fig.~4) & 4.86 & 21.7 & 4.468 & 5.15 & 23.0 & 4.474 & $-0.13\%$ \\
Fit, 1--4\,ns & 4.88 & 22.1 & 4.538 & 5.16 & 23.2 & 4.505 & $+0.75\%$ \\
Fit, 1.2--4\,ns & 4.87 & 22.2 & 4.558 & 5.15 & 23.2 & 4.499 & $+1.33\%$ \\
Fit, 3--4\,ns & 4.88 & 21.7 & 4.445 & 5.17 & 22.9 & 4.437 & $+0.17\%$ \\
100\,ps blocks after 1.2\,ns ($n=28$) & & & $4.559\pm0.062$ & & & $4.465\pm0.057$ & $+2.10\%$ \\
400\,ps blocks, 2--4\,ns ($n=5$) & & & $4.482\pm0.061$ & & & $4.442\pm0.037$ & $+0.88\%$ \\
\midrule
$\nabla T$ in Si / Ge, 2--4\,ns & \multicolumn{3}{c}{$-0.100$ / $-0.543$} & \multicolumn{3}{c}{$-0.131$ / $-0.469$} & \\
$J/|\nabla T|$ in Si / Ge (W\,m$^{-1}$\,K$^{-1}$)$^{a}$ & \multicolumn{3}{c}{48.8 / 9.0} & \multicolumn{3}{c}{39.4 / 11.0} & \\
\bottomrule
\end{tabular}
\end{table}

\begin{table}[htbp]
\centering
\caption{Phonons of the 24-atom Si$_6$Ge$_6$ superlattice (frequencies in
THz). MAE is the mean absolute deviation from DFT over all 72 bands along
$\Gamma$--X--M--$\Gamma$--Z--R--A--Z. DOS overlap is
$\int\min(g,g_{\mathrm{DFT}})\,\dd\nu/\int g_{\mathrm{DFT}}\,\dd\nu$. $^{a}$Located near $\Gamma$ (DFT), along $\Gamma$--X (CPL) and at $\Gamma$ (PARS).}
\label{tab:s-phonon}
\begin{tabular}{@{}p{5.2cm}p{1.8cm}p{1.8cm}p{2.4cm}@{}}
\toprule
\textbf{Quantity} & \textbf{DFT} & \textbf{CPL} & \shortstack{\textbf{PARS (NEP}\\\textbf{distillation)}} \\
\midrule
Path MAE vs DFT & -- & 0.18 & 0.23 \\
Path MAE, modes below 10\,THz & -- & 0.19 & 0.17 \\
Path MAE, modes above 10\,THz & -- & 0.17 & 0.38 \\
Mean shift of modes above 10\,THz & -- & $-0.15$ & $-0.38$ \\
Highest $\Gamma$ frequency & 14.94 & 14.64 & 13.93 \\
Lowest frequency on path$^{a}$ & $-0.20$ & $-0.31$ & $-0.04$ \\
DOS overlap with DFT & -- & 0.82 & 0.82 \\
\bottomrule
\end{tabular}
\end{table}

\FloatBarrier
\section*{Supplementary Note 8: Prepared-embedding time and memory (Supplementary Fig.~S2)}
\label{supnote:scaling}

Supplementary Fig.~S2 compares PARS with our implementation
of the published IDEAL embedding procedure\cite{kong2023ideal}
for one or eight centres at $N=1{,}280$--$100{,}000{,}000$. Parent cells are
periodic copies of a retained Si/Ge MD frame with the same ordered, previously
verified source OOD environments, not new large-cell MD or whole-frame OOD
tests. Both methods protect a 5~\AA{} core and cap outputs at 128 atoms.
Runs use one CPU computation thread on Kunpeng 920~7270E, a 128\,GiB
allocation and a 112\,GiB address-space limit; nodes are not exclusive.
Sizes up to $1{,}024{,}000$ atoms have three fresh-process repetitions;
larger sizes have one matched-node measurement per method/count.
At $N=1{,}280$, $10{,}240$ and $20{,}480$, same-core interleaved reruns use a
benchmark-only PARS periodic-KDTree distance-threshold dispatch with native
fallback; structure fingerprints and acceptance are unchanged. Other sizes
retain measured references using the unchanged large-system preparation path.

Time covers the complete batch call after frame-specific spatial data,
references and runtimes are ready. All PARS construction, physical gates
and scoring, and IDEAL's 100 actual SOAP-scored boundaries and up to 200
peripheral-optimization steps are retained; the step cap is not proof of
convergence. Frame preparation, common OOD inference, post-embedding FIRE repair, IO and external acceptance are excluded. Peak RSS covers process startup through embedding
return, including input, libraries, references and preparation temporaries,
and is captured before external acceptance.

Spatial indexing confines PARS candidate work to a local window scored by
Eq.~\eqref{eq:s-physical-score}; the tested IDEAL implementation also
repeats full-parent scans and evaluates SOAP coordinate derivatives.
At $N=1{,}024{,}000$, both single-centre outputs contain 50 atoms and pass
the common geometric checks: PARS takes $0.104$\,s versus $20.514$\,s for
IDEAL (Table~\ref{tab:s-scaling}). The IDEAL median-time repetition spends
$3.464$\,s in boundary search and $17.007$\,s in optimization.
This is an implementation-level cost explanation, not a constant-time law:
PARS index preparation and memory depend on $N$, and IDEAL could also use
local indexing.

Unrepaired outputs are audited for centre/species identity, complete
core-neighbour sets, periodic/rigid core geometry ($10^{-6}$~\AA{}
tolerance), common parent-aware distance floors and finite positive cells.
With $C_{\mathrm{pass}}=T_{\mathrm{emb}}/n_{\mathrm{pass}}$, completed
eight-centre batches pass $8/8$ for PARS and $5/8$ for IDEAL; at
$N=1{,}024{,}000$, their costs are $0.0973$ and $39.1825$\,s per accepted
structure, with mean sizes $63.625$ and $56.5$ atoms. No repair is used.
IDEAL's ten configurations at $N\ge8{,}192{,}000$ have preparation
\texttt{MemoryError}s under this budget, not valid prepared-call time/RSS.
The fits $y=c+a(N/10^6)^b$ describe completed measurements only; no failed
region is extrapolated or complexity inferred. 
Geometric acceptance does not establish electronic-structure equivalence.

\makeatletter
\setlength{\@fpsep}{12pt}
\makeatother
\begin{figure}[!htbp]
\centering
\includegraphics[width=\textwidth]{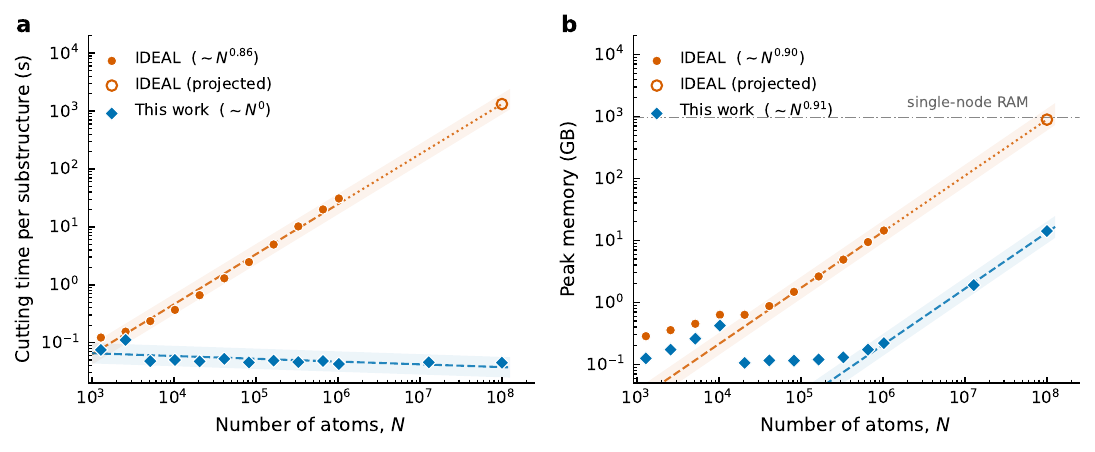}
\caption{\textbf{Prepared-embedding time and peak-memory scaling.}
\textbf{a,b}, Batch wall time for one and eight centres after frame
preparation. \textbf{c,d}, Process peak RSS, including frame-preparation
allocations (Supplementary Note~8). Filled markers show medians of three
fresh-process repetitions with minimum--maximum error bars; open markers
show single measurements above $1{,}024{,}000$ atoms. Dashed curves are
descriptive fits within the completed measured range, not complexity laws.
Crosses below the axes denote frame-preparation \texttt{MemoryError} under
the 128\,GiB allocation/112\,GiB address-space budget; their vertical
positions are categorical, not time or memory values. Acceptance and
actual output sizes are reported separately; no electronic-fidelity
equivalence is implied.}
\label{fig:s-scaling}
\end{figure}

\begin{table}[!htbp]
\centering
\caption{Representative measured batch times and peak
process RSS for Supplementary Fig.~S2.
The $1{,}024{,}000$-atom entries are three-repetition
medians; larger systems are single measurements.
Dashes indicate a frame-preparation
\texttt{MemoryError}, not zero cost.
The IDEAL results refer to our implementation of the
published embedding procedure.}
\label{tab:s-scaling}
\footnotesize
\setlength{\tabcolsep}{3.2pt}
\begin{tabular}{@{}lcrrrr@{}}
\toprule
$N$ & Centres & \shortstack{PARS\\time (s)} & \shortstack{IDEAL\\time (s)} &
\shortstack{PARS\\RSS (GiB)} & \shortstack{IDEAL\\RSS (GiB)} \\
\midrule
$1.024\times10^6$ & 1 & 0.104 & 20.514 & 0.337 & 25.419 \\
 & 8 & 0.778 & 195.912 & 0.336 & 25.419 \\
\addlinespace[2pt]
$4.096\times10^6$ & 1 & 0.113 & 28.436 & 0.838 & 98.691 \\
 & 8 & 0.823 & 246.871 & 0.838 & 98.686 \\
\addlinespace[2pt]
$1.0\times10^8$ & 1 & 1.017 & -- & 16.422 & -- \\
 & 8 & 2.273 & -- & 16.423 & -- \\
\bottomrule
\end{tabular}
\end{table}
\FloatBarrier

\clearpage
\section*{Supplementary References}
\begingroup
\renewcommand{\section}[2]{}%
\bibliography{fieldmat_pars_references}
\endgroup